# Avatar Exposure and Strategic Coordination in Virtual Reality: Evidence from a Threshold Public Goods Experiment

Manuela Chessa[1*], Michela Chessa[2,3†], Lorenzo Gerini[1‡],
Matteo Martini[1§], Kaloyana Naneva[2¶] Fabio Solari[1‖]

[1]University of Genoa, Italy
[2]Université Côte d'Azur, GREDEG, CNRS, France
[3]INRIA, Université Côte d'Azur, France

**Abstract**

Digital platforms increasingly support collective action initiatives, yet coordinating geographically dispersed users through digital interfaces remains challenging, particularly in threshold settings where success requires critical mass participation. This study investigates how avatar-based social representation in Virtual Reality (VR) influences coordination in threshold collective action problems. Through a randomized controlled experiment with 188 participants organized in 94 pairs, we examine whether brief avatar exposure affects perceived co-presence and coordination outcomes in a two-player threshold public goods game implemented as a real-effort recycling task. We manipulate a single design feature: participants either briefly interact through avatars before the main task (Pre-Task Avatar treatment) or complete an equivalent activity individually without peer visibility (No Pre-Task Avatar treatment). Our findings reveal that minimal avatar exposure significantly increases perceived co-presence and improves strategic coordination. Participants exposed to peer avatars achieve higher social welfare by coordinating to avoid wasteful over-contribution beyond the threshold. Additionally, we identify VR presence—the sense of "being there" in the virtual environment—as a stronger predictor of task performance than co-presence itself. This research contributes to Information Systems theory by establishing causal pathways from specific design features to presence to coordination outcomes, demonstrates VR as a rigorous experimental methodology for IS research, and provides actionable insights for designing collaborative platforms supporting sustainability initiatives and threshold collective action problems.

[*]manuela.chessa@unige.it, https://orcid.org/0000-0003-3098-5894
[†]Corresponding author, michela.chessa@univ-cotedazur.fr, https://orcid.org/0000-0003-4218-9369
[‡]lorenzo.gerini@edu.unige.it, https://orcid.org/0009-0000-8988-6909
[§]matteo.martini@edu.unige.it, https://orcid.org/0009-0006-3929-5055
[¶]kaloyana.naneva@univ-cotedazur.fr, https://orcid.org/0009-0001-7299-4249
[‖]fabio.solari@unige.it, https://orcid.org/0000-0002-8111-0409

# 1. Introduction

Digital platforms have an increasingly important role in the transformation of industries and even in reshaping entire economies (de Reuver et al., 2018). They have also become central to the organization of collective action across diverse domains, where individuals contribute time and resources to goals that extend beyond their personal interest. Examples include encouraging activities like crowdfunding (Zhang et al., 2020), open source software development (Oreg & Nov, 2008; Zhao & Deek, 2004), circular economy (Kovacic et al., 2020; Tian et al., 2024), and creative thinking (Karakaya & Demirkan, 2015). Platforms such as Kickstarter and Wikipedia depend on the coordinated efforts of geographically dispersed users who may never meet in person yet still need to cooperate toward the achievement of shared objectives (Modaresnezhad et al., 2020). However, a persistent challenge these platforms face is the difficulty of fostering effective collaboration in online communities when users interact through digital interfaces rather than face-to-face (Faraj et al., 2011). This limitation becomes especially problematic in threshold collective action settings, where success depends on a critical mass of contributions–if too few participate, everyone's effort is lost.

Consider, for example, the growing body of literature highlighting the increasing relationship between digital transformation and sustainable development (H. Li et al., 2025). Communities worldwide have deployed mobile applications and web platforms to encourage household recycling and waste reduction. However, these platforms encounter the fundamental aforementioned coordination challenge: individual households must together reach a level of participation high enough for the digital recycling system to function effectively, yet each household makes decisions independently, often with limited awareness of others' actions. The transition toward a circular economy–one of the most urgent sustainability challenges of our time–relies heavily on such coordinated collective action at the residential level. Understanding how to design digital systems in order to facilitate coordination in collective action problems represents a critical challenge for Information Systems (IS) research and practice (Zeiss et al., 2021).

Among digital platforms, recent advances in Virtual Reality (VR) offer new possibilities to address coordination challenges in such remote collaborative environments. Unlike the more traditional web or mobile interfaces, VR systems can convey a wide range of social cues through visual, audio and haptic information. VR uniquely enables users to experience a realistic environment in which they are embodied through their own avatar while perceiving and interacting with others' avatar representations. From a design perspective, these capabilities enable VR platforms to recreate elements provided by face-to-face interaction, such as peer visibility and non-verbal communication, that may facilitate coordination (Dzardanova et al., 2024; Herrera et al., 2020). In research, VR provides unprecedented opportunities for establishing controlled experimentation: it allows for the systematic manipulation of social cues that are often impossible to isolate in conventional laboratory or field studies. Together, these qualities position VR as both a promising design space for next-generation collaborative platforms and an ideal methodology for rigorously investigating how specific system features influence user behavior, while maintaining experimental control and the ecological validity of authentic social interactions.

In this paper, we investigate how avatar-based social cues in VR influence cooperation in a threshold public goods game, providing experimental evidence on how immersive environments can support coordinated collective action. Yet fundamental questions remain about how specific design features influence collaboration in collective action contexts. Should platforms invest in sophisticated avatar systems to represent users to each other? Do these social representation features genuinely improve coordination? Through what mechanisms might design choices influence user decisions? Moreover, how do individual differences in technology experience moderate these effects? Addressing these questions requires understanding how social presence operates in collaborative virtual systems.

To fill these gaps, more specifically, we examine: *(i)* Whether brief exposure to peer avatar representation increases perceived co-presence in collaborative VR environments and *(ii)* Whether peer representation improves contribution and coordination success in threshold collective action tasks. Additionally, recognizing that real-world collective action often occurs under temporal constraints, we examine *(iii)* How time availability influences contribution choices and accuracy.

We address these questions through a randomized controlled experiment in which 188 participants, organized into 94 pairs, performed a coordinated recycling task in an immersive environment using VR headsets. The task is implemented as a two-player threshold public goods game requiring strategic coordination in a real-effort task. In this setting, each participant chooses for each object whether to recycle it (contributing to the public good) or dispose of it in the black bin as unsorted waste (not contributing, but reducing own recycling effort). Total contributions of players toward the public good must reach a specified minimum level for the group reward to be obtained; below that threshold, the group reward is not provided. We manipulated a single design feature: in the *Pre-Task Avatar* treatment, pairs briefly met and interacted through avatar representations in a shared virtual space before the main task; in the *No Pre-Task Avatar* treatment, participants completed an equivalent preliminary activity individually without peer visibility. This minimal treatment allows us to isolate the causal effect of social representation design on subsequent coordination behavior while holding constant all other system features, task characteristics, and environmental factors.

Our choice of a recycling task serves multiple purposes aligned with IS research priorities. First, it provides a framing that mirrors real-world sustainability platforms where coordination among distributed users is essential. Second, the environmental context connects to important practical applications, because understanding how users coordinate is central to the success of digital platforms supporting sustainability initiatives. Third, implementing recycling as a real-effort task requiring identification of material types, separation of components, and correct placement under decreasing time availability enables us to study how the investigated design feature influences both strategic choices and task execution quality.

We employed experimental methods to address our research questions. Experimental approaches have become a well-established and valued methodology within IS scholarship (Ballatore et al., 2020). IS experiments–conducted both in laboratory (Hashim et al., 2017) and field settings (Benbasat & Zmud, 1999)–have been extensively validated as rigorous tools for theory testing (Dennis & Valacich, 2001). These methods are particularly effective for bridging the gap between theoretical models and actual user behavior, enabling researchers to systematically examine how individuals acquire and process information, and how they create value through decision-making (Gupta et al., 2018).

We measured multiple dimensions of user experience and behavior. Co-presence, referring to the sense of being together with another person in a shared space (i.e., first-order social presence) was assessed using the Networked Minds Social Presence Inventory (Biocca & Harms, 2003). Physical presence within the VR environment was measured using both the Igroup Presence Questionnaire (Schubert et al., 2001) and selected items from the revised Witmer-Singer Presence Questionnaire (Witmer & Singer, 1998). We also assessed pro-environmental attitudes using an adapted recycling behavior questionnaire (Czajkowski et al., 2017) and collected demographic information including VR familiarity. Behavioral measures included number of correctly recycled items, incorrectly recycled items, items disposed of as general waste, and temporal patterns of choices.

Our experimental findings reveal important insights about how social representation features influence coordination in collaborative systems. We find strong evidence that brief avatar exposure significantly increases perceived co-presence, demonstrating that minimal design features can effectively create social presence. However, avatar representation does not directly increase individual contribution quantity; instead, it improves strategic coordination. Participants in the *Pre-Task Avatar* treatment placed more items in the black bin (reducing recycling effort) yet achieved significantly higher social welfare, indicating better coordination to avoid wasteful over-contribution beyond the threshold. This finding illustrates a critical distinction often overlooked in collaborative platform research: more contribution does not always equal better outcomes in threshold scenarios. Additionally, we identify VR presence—users' sense of "being there" in the virtual environment (VE)–as a more robust predictor of task performance than co-presence itself, suggesting that investments in immersion quality may be as important as investments in social features for certain collaborative tasks.

Our research makes several contributions to IS theory and practice. Firstly, we demonstrate how specific design features create presence in collaborative systems and reveal the causal pathway from design to presence to coordination outcomes, addressing a critical gap in prior research that examined presence as an emergent property rather than investigating which design choices create it. Secondly, methodologically we establish VR as a rigorous experimental approach for IS research, showing how researchers can isolate design effects while maintaining ecological validity through real-effort, behaviorally-consequential tasks. Thirdly, we provide actionable insights for platform designers as our findings suggest that even minimal social representation (brief, non-verbal avatar interaction) can influence coordination, though the mechanism operates through strategic adjustment rather than increased effort. Finally, we contribute to understanding coordination in threshold collective action problems, a context increasingly relevant as digital platforms support crowdfunding campaigns, community initiatives, and sustainability programs where collective goals require the participation of a critical mass.

The remainder of this paper proceeds as follows. Section 2 reviews the relevant literature on social presence in virtual environments. Section 3 provides the experimental setup and procedures, detailing the VR system implementation, training protocols, treatment manipulation, and measurement instruments. Section 4 presents our experimental hypotheses. Section 5 reports our experimental results. Section 6 discusses implications of our findings, future research directions and finally concludes with a synthesis of contributions and recommendations for practitioners and researchers.

## 2. Related Literature

In the domain of virtual environments, the sense of presence has been formerly conceptualized as "the sense of being there", i.e., a situation during which people behave in a VE in a way that is similar to what their behaviour would have been in a similar real-life situation ( Slater, 2002). Then, the concept of presence has been formalized as a multidimensional construct that comprises physical, social, and self-presence (Lee, 2004). Existing IS research has further established that social presence can strongly influence trust, satisfaction, and behavioral outcomes within digital and virtual settings (Gefen & Straub, 2003; Hess et al., 2009; Qiu & Benbasat, 2009; Robert et al., 2009). In collaborative systems, social presence is believed to strengthen prosocial motivation by making the presence of real collaborators more salient. Particularly relevant to collaborative platforms is *co-presence*—the degree to which users feel that they are together in the same space with others (Biocca et al., 2003)—which represents the foundational level of social presence necessary for collaborative interaction. Prior evidence suggests that behavioral realism (e.g., gaze, movement, and proxemic behavior) consistently contributes to greater levels of co-presence (Von der Pütten et al., 2010), whereas photographic and anthropomorphic realism alone result in mixed effects (Nowak & Biocca, 2003). Moreover, perceived agency of virtual social interactions also matters: users report stronger levels of social presence when interacting with human-controlled avatars as opposed to algorithmic agents (Fox et al., 2015). These findings indicate that design choices concerning digital representation may influence users' perceptions of collaborative partners and, consequently, coordination behavior.

However, two critical gaps limit our ability to translate existing research into actionable design principles for collaborative platforms. First, most studies treat social presence as an *outcome* of sustained, communication-rich interaction, rather than examining the specific system design features that are responsible for its *creation* (Khalifa & Shen, 2004). For platform designers facing resource allocation decisions, knowing that presence matters is insufficient; they need to understand which specific design features result in higher social presence and lead to better coordination outcomes.

Early theories described social presence as a property of the medium itself, assuming that richer communication channels inherently produced stronger presence (Short et al., 1976). However, subsequent research has shown that presence is not determined solely by media richness—text-based forms of computer-mediated communication can also elicit strong feelings of social presence (Oviedo & Tree, 2024). 3D virtual environments provide unique conditions for *being together*, allowing for different communication possibilities (Sivunen & Nordbäck, 2015). Yet, regardless of the well-studied fact that presence matters in online collaboration (Boughzala et al., 2012; Srivastava & Chandra, 2018), we still know little about which specific interface features can generate a sense of presence in VR—independent of sustained communication among users. Our study contributes by experimentally isolating one such element of interaction through the study of the role of digital representations.

Second, there are studies on collaborative virtual environments within what is known as avatar-mediated communication. This research has focused on communication-rich collaboration, where participants interact through synchronous verbal or text exchanges supported by non-verbal cues such as gestures or gaze (e.g., (Aburumman et al., 2022); (C. Li et al., 2025)). Far less attention has been given to minimal, non-verbal forms of social exposure, such as brief visual co-presence through avatar representation, where users share the same virtual space without direct communication. Understanding whether such limited interaction can still evoke a sense of connection sufficient to support coordination is particularly relevant for contexts characterized by brief or low-communication encounters, such as contribution-based or event-driven collective action initiatives taking place within VR environments.

# 3. Experimental Setup and Procedures

The experiment was conducted across 94 sessions. Participants were recruited via word of mouth, social media posts, and the distribution of flyers. At the beginning of each session, participants recruited in pairs were welcomed by the experimenters in two different locations to prevent them from meeting in person. They were then taken to separate rooms, asked to sign a consent form, and guided through reading the general instructions by the experimenters. Participants were asked whether they wanted to perform the experiment in English or in Italian, and all materials (instructions in Appendix B.1, voice-over, screen information) were provided accordingly. They were then assisted in putting on the head-mounted VR equipment, and the VR experiment began.

The VR experiment comprised four distinct parts[7]. At the beginning of each part, a voice-over, together with additional screens on top of the VR environment provided specific instructions and guidance for the respective part of the experiment. In the first part, participants completed a training session to ensure a uniform baseline skill level and familiarity with the VR environment (see Section 3.1). In the second part, participants were virtually transferred to a 3D reconstruction of the Italian coastal town of Positano, where they performed a preliminary sorting task either alone or together with their matched partner, depending on the treatment (see Section 3.2). In the third part, participants were guided by the voice-over through reading on the screen the specific instructions of the main recycling task (see Appendix B.2), and then had five minutes to complete it (see Section 3.3). In the final fourth part, once the task was completed, participants returned to the Positano environment to observe the consequences of their collective effort and learn about their resulting monetary gain (see Section 3.4).

Upon completion of the VR session, participants were then helped to remove the VR headset and invited to complete a final questionnaire (see Section 3.5). Finally, they were paid the corresponding amount earned during the experiment, plus a €5 show-up fee.

## 3.1. Virtual Reality setup material and familiarization

The VR environment was developed using Unity and designed to support collaborative interactions in a controlled setting. Participants experienced the environment through a Meta Quest 3 head-mounted display (HMD), paired with hand-held Touch Plus controllers that allowed to manipulate objects and interact naturally with the virtual space, with raycast modality. The system was built to support networking and multiplayer functionalities, enabling experiments to quickly create or join sessions for the collaborative experience. Further details about the app development can be found in Appendix C.

To ensure that all participants were familiar with the VR interface before beginning the main task, a training session was conducted. Participants entered an abstract virtual space, where they were instructed on how to use the hand-held controllers. Each stage of the training included a checkpoint that had to be completed before participants could proceed further. Once each sub-task was completed, the experimenters could suggest that participants repeat it, if they were not confident that the concept had been fully learned. When a sufficient level of familiarity had been achieved, the procedure moved on to the subsequent sub-task.

---

[7]The collaborative virtual environment has been presented at the 2025 IEEE Conference on Virtual Reality and 3D User Interfaces. (Chessa et al., 2025).

First, participants were instructed on how to move through the virtual space. Since the main task involved placing items in bins, they were then taught how to handle VR objects using the hand-held controllers. A virtual cube appeared in front of them, which they had to manipulate through different actions. Participants were first asked to hold the cube with the right controller and then with the left, and move it from side to side until it was placed in a designated space. Next, they were asked to move the cube closer to and farther away from themselves. After completing these exercises, participants were presented with labeled cubes and were taught how to examine the cubes from all sides while holding them. They then had to sort the cubes according to their labels. Finally, they were asked to separate pairs and triplets of cubes attached together, and turn them to identify their labels in order to place them in the corresponding space.

Once the training task was completed, participants could choose their avatar from a selection of 30 predefined options. They were diversified in terms of style, gender, ethnicity, and age, but not customisable. Participants could view their avatar's upper-body virtual reflection including head and hands in a mirror within the environment before confirming their choice. Once confirmed, they were teleported to the scene for the next phase.

### 3.2. Preliminary task

The preliminary task was conducted in a virtual reconstruction of the Italian coastal town of Positano. Participants were able to explore and enjoy the beautiful environment. During their virtual visit, they were guided with arrows to reach a specific spot where they were invited to perform a simple sorting task. The preliminary task lasted until the sorting task was completed. Participants were presented with a pile of objects consisting of shells and fish, which they had to sort into two corresponding bins.

Our two treatments differed only in this preliminary task. In the *No Pre-Task Avatar* treatment, each participant explored the town of Positano and performed the simple sorting task individually, sorting their own pile of objects without seeing the other participant (see Figure 1a). In the *Pre-Task Avatar* treatment, the two participants in a pair shared the same virtual space and were able to perform the sorting task of a shared pile of objects together (see Figure 1b). No verbal communication was allowed. Participants were encouraged by the guiding voice-over to exchange a greeting gesture upon meeting and could interact non-verbally through gestures if they wished. After completion of the shell and fish sorting task, the participants were subsequently teleported to the next recycling task.

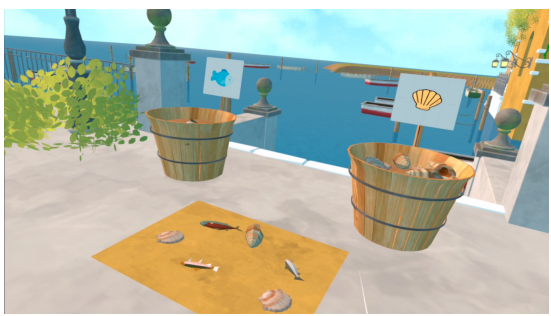

(a) The sorting baskets and the objects to be sorted on the floor in a *No Pre-Task Avatar* treatment.

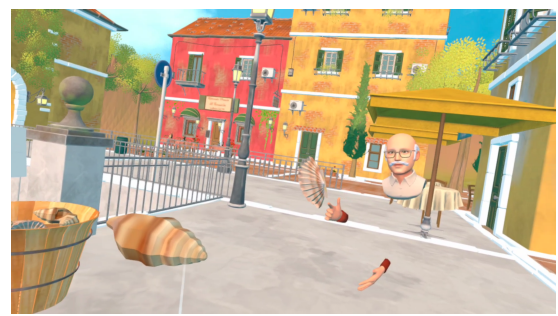

(b) The preliminary task being performed in a *Pre-Task Avatar* treatment

Figure 1: Preliminary task in Positano

## 3.3. The recycling task

**A 2-player threshold public goods game**

In our main task, a recycling task, we adopted a 2-player threshold public goods game framework, in which group benefits are realized only if a minimum group contribution is reached. This structure closely mirrors environmental behaviors such as recycling, where outcomes are meaningful only if a critical mass of participants engages in the activity.

In the game, $N = \{1, 2\}$ is the *set of players*, which we call participants. Each participant has an initial endowment of $M = 56$ items to be sorted into different bins. Both participants have the same set of items and perform the recycling task simultaneously and independently.

A strategy for player $i$ is given by a vector $x_i = (x_i^B, x_i^R, x_i^W)$, with $x_i^B + x_i^R + x_i^W \leq M$, where:

- $x_i^B \in \{0, M\}$ is the number of items participant $i$ puts in the black bin. Sorting an item into the black bin yields a reward of €0.20;

- $x_i^R \in \{0, M\}$ is the number of items participant $i$ correctly recycles. Correctly recycling an item yields €0.10;

- $x_i^W \in \{0, M\}$ is the number of items participants $i$ incorrectly recycles, meaning, the number of items placed into the wrong recycling bin. Incorrectly recycling an item yields €0.

We define $x_i^{\text{Thrown}} = x_i^B + x_i^R + x_i^W \in \{0, M\}$ the number of items participant $i$ throws into any bin, and $X = x_i^R + x_{-i}^R \in \{0, 2M\}$ the total number of correctly recycled items by a pair (where $x_{-i}$ denotes the number of correctly recycled items by participant $i$'s peer). The threshold is given by $T = 84$ (75% of all items combined for both participants) and when the pair correctly recycles at least T items, then each participant receives a bonus of €10. A further €3 bonus is awarded if a participant disposes of all her 56 items ("room cleared"). This additional bonus is introduced to simulate the fact that the primary goal – in the experiment as in daily life – remains to eliminate all the trash in the room.

Formally, the game payoff for participant $i \in N$ is defined as follows:

$$\pi_i\left(x_i^B, x_i^W, x_i^R, x_{-i}^R\right) = 0.20 \cdot x_i^B + 0.10 \cdot x_i^R + \begin{cases} 10, & \text{if } X \geq T \\ 0, & \text{otherwise} \end{cases} + \begin{cases} 3, & \text{if } x_i^{\text{Thrown}} = M \\ 0, & \text{otherwise.} \end{cases}$$

Before proceeding with our analysis, it is worth noting the following observations:

**Observation 1.** It is always Pareto optimal to choose a strategy such that $x_i^B + x_i^R = 56$, as it is never convenient to deliberately choose not to place an item or to place an item in the incorrect bin, since doing so yields no individual reward.

**Observation 2.** The maximum possible payoff for a participant $i \in N$ is €26.40, which can be obtained when using what we call a "partial free-riding strategy". That is, a participant relies as much as possible on the recycling effort of their peer. In our game, it means that when the threshold $T = 84$ is met, participant $i$ minimizes their own recycling effort by recycling only 28 items and placing the remaining 28 items in the black bin, while the peer recycles all 56 of their own items correctly. In such a situation, their peer gets a final payoff of €18.60.

**Observation 3.** Our game is a coordination problem and it also exhibits two types of pure strategy Nash equilibria. The first is the symmetric pair of strategies in which both participants fully free-ride and place all their items in the black bin, resulting in no provision of the public good and an individual payoff of €14.20. In this situation, if a participant expects the other not to contribute to the recycling task, any positive contribution is wasteful because it is impossible to meet the threshold alone. The second type of pure strategy Nash equilibria occurs when participants' contributions meet exactly the threshold: any individual deviation is not profitable, as contributing less causes the threshold to be missed, while contributing more merely reduces one's own payoff without increasing the collective return. In threshold public goods games, in fact, contributing beyond the threshold yields no additional benefits and represents an inefficient use of resources. There are many such equilibria (including the one presented in Observation 2), since any pattern of contributions that sums exactly to the threshold qualifies, which highlights the complexity of coordinating. Only one of these Nash equilibria is symmetric: the one where each participant contributes the same amount and correctly recycles 42 items, while placing the remaining 14 items in the black bin. The resulting payoff of both participants is equal to €20. However, this equilibrium does not maximize social welfare (when we define *social welfare* as the sum of a pair utilities).

**The VR real-effort implementation**

Through Virtual Reality, our game was implemented as a real effort task. Importantly, in both the *No Pre-Task Avatar* and *Pre-Task Avatar* treatments, the peer's representation and actions were not visible while performing the recycling task. Participants each entered a virtual room containing a very cluttered desk with various objects present both on the desk surface and on the floor, and were asked to clean the room. The different bins corresponding to the object materials (organic, e-waste, plastic, metal, paper, general waste, glass), were arranged in front of the desk, with the black bin positioned centrally and the recycling bins arranged around it (see Figure 2a). The task consisted of sorting the objects by material type (see Figure 2b) and placing each item in either the black bin or the corresponding recycling bin. When participants were uncertain about the correct recycling bin, material type labels were available to help them make the right decision.

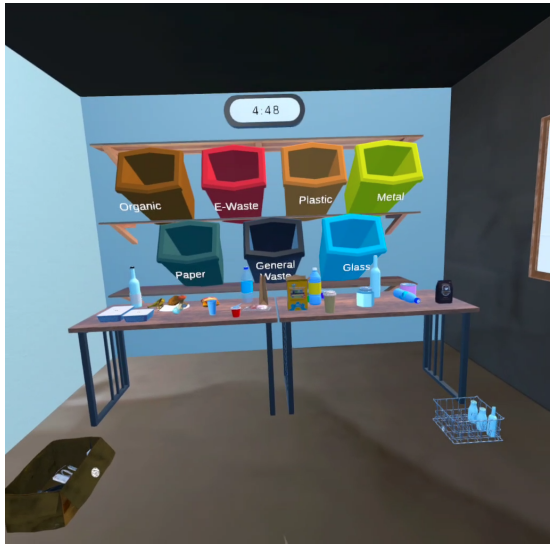 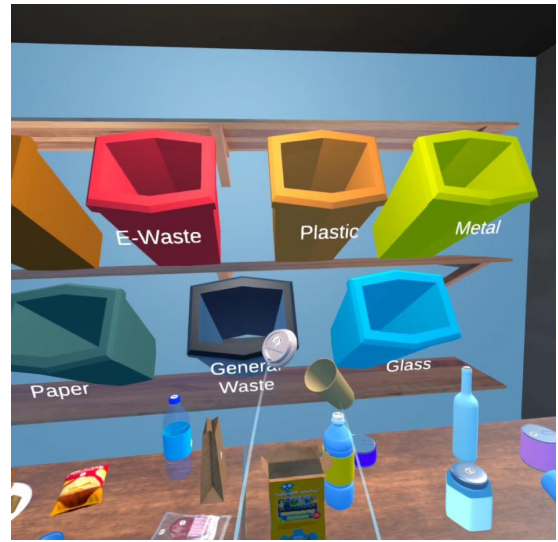

(a) The virtual room to be cleaned    (b) Material separation during the task. Some material labels are visible on the objects

Figure 2: Recycling task at home

Several factors could influence the final outcome of the game, regardless of the initial strategy chosen by a player and the theoretical observations 1-3. First, participants were given a 5-minute time limit to complete the task. Time pressure was induced by displaying the remaining time on a clock positioned on top of the desk (see Figure 2a), with a voice-over announcement when only 1 minute remained. This time constraint was imposed to make the task more challenging while simulating real-world situations where people prefer not to spend excessive time disposing of trash. Consequently, the decision not to place an item in a bin, even if suboptimal, might have resulted from time pressure. This constraint could also have influenced decisions to throw items in the black bin, as players might find it more convenient to complete the task in the easiest and fastest way possible as time runs out.

Second, correctly recycling items could be challenging. Some objects consisted of multiple components (e.g., a bottle and its cap) that needed separation before disposal (see Figure 2b). Then, identifying the correct bin was not straightforward (e.g., for electronic items), or participants could miss the labels and make incorrect choices when rushed. These factors could lead to unintentional errors and influence the recycling task outcome.

Third, strategic considerations regarding the peer's behavior could also be affected by these factors, as participants might be confident in their own abilities while doubting their partner's capability to perform the task correctly and on time.

### 3.4. Outcome revelation and environmental feedback

Once the task was completed, participants were returned to the virtual Positano environment to observe the consequences of their collective recycling effort. They were presented with either a utopian version of the town (see Figure 3a) – displayed when the recycling threshold had been met – or a dystopian version (see Figure 3b) when the threshold was not reached. During this revelation phase, both the performance outcomes and monetary rewards for each participant and their peer were disclosed.

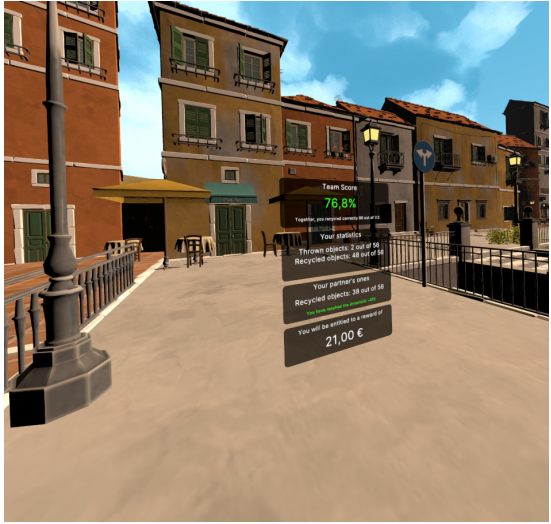 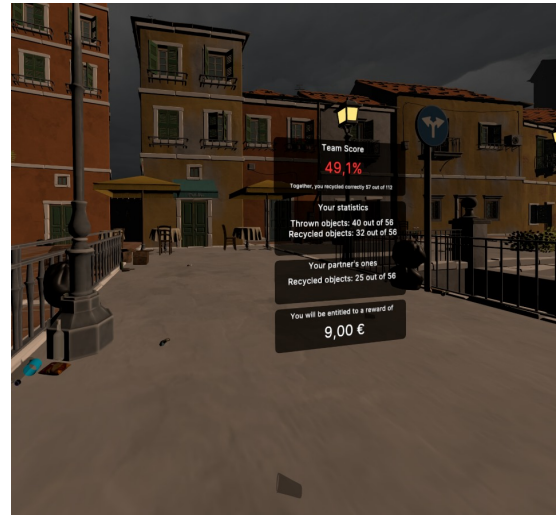

(a) Pair threshold reached  (b) Pair threshold not reached

Figure 3: Environmental feedback after the recycling task

### 3.5. Questionnaires

Following the recycling task, participants completed a post-experimental questionnaire on a tablet. The questionnaire included standardized measures and custom items to assess variables relevant to our research questions.

**Demographics.** The questionnaire began with demographic questions. These were used in our study to define the variables *Gender, Age, Nationality, Level of completed studies, Current (studying/working) situation, VR use frequency* and *Risk preferences*. Additionally, we included a *Language* variable to distinguish between participants who requested to complete the experiment in Italian versus those who opted for English.

**Pro-environmental behavior assessment.** We measured participants' real-life recycling behavior using an adapted version of the questionnaire developed by Czajkowski et al. (2017). This instrument captures key determinants of recycling behavior, including social pressure, moral motivation, and perceived private costs or effort. Based on the responses provided in this part of the questionnaire, we defined an *Environmental score* for each participant.

**VR presence measures.** To assess participants' sense of presence in the virtual environment, we employed two complementary standardized instruments. The "Igroup Presence Questionnaire" (*IPQ*) by Schubert et al. (2001) includes 14 items across four dimensions: general presence, spatial presence, involvement, and experienced realism. Based on the responses provided in this part of the questionnaire, we computed an index for each dimension and for each participant ($G$ = General Presence, $SP$ = Spatial Presence, $INV$ = Involvement, and $REAL$ = Experienced Realism) and a general *IPQ* index for each participant. Additionally, we selected 10 items from the revised version of the "Presence Questionnaire" by Witmer and Singer (1998) (*WS*) revised by the UQO Cyberpsychology Lab, focusing on aspects not covered by the IPQ, particularly navigation and object manipulation in VR environments. Based on the responses provided in this part of the questionnaire, we defined a *WS* index for each participant.

**Sense of co-presence assessment.** To measure whether participants felt their peer was genuinely present

"with them" in the virtual space, we used the co-presence subscale from the "Networked Minds Social Presence Inventory" (Biocca & Harms, 2003). This 8-item measure captures first-order social presence, assessing the degree to which users feel they are sharing the same virtual space with another person. The questionnaire is composed of two sub-categories: perception of self, and perception of the other. We defined the two variables *Perception self* and *Perception other*, together with an averaged score based on all items, that we call *Co-presence* index, for each participant.

**Recycling motivations.** We included seven custom items using 5-point Likert scales to understand participants' underlying motivations for their recycling decisions during the task, providing insight into their decision-making processes. Based on the responses provided in this part of the questionnaire, we then defined the following variables: *Quantity, Quality, Collective good, Private good, Reciprocity, Competition* and *Cooperation* (see Appendix A.5).

**Control measures.** Finally, we assessed risk attitudes and prior VR experience.[8] We defined the variables *Risk preferences* and *VR use frequency*.

Complete questionnaire items for all measures, along with information on how the different indices and scores were computed, are provided in Appendix A.

## 4. Experimental Hypotheses

The experiment included two treatments: one in which participants met and shortly interacted with their peer represented by an avatar (*Pre-Task Avatar*) prior to performing the main recycling task, and one in which they did not (*No Pre-Task Avatar*). In both treatments, participants met their peer avatar at the end of the experiment, and viewed the final results of the main task together.

This design isolates the effect of peer visibility on social interaction and coordination when performing an experimental task in a high-immersion VR environment. In this section, we present the experimental hypotheses that guided our experimental research.[9].

Previous studies on virtual environments have demonstrated that social presence is typically higher when a visible peer representation is available in settings with verbal communication (Nowak and Biocca, 2003). Our study does not involve any verbal communication; therefore, we focus solely on the first degree of social presence: co-presence. In the *No Pre-Task Avatar* condition, the two peers did not share the initial immersive environment. In the *Pre-Task Avatar* treatment condition, instead, peers met and even performed a small preliminary activity together. This drove the formulation of our first hypothesis:

- **H0**: The sense of co-presence is significantly higher in the *Pre-Task Avatar* treatment than in the *No Pre-Task Avatar* treatment.

The *Pre-Task Avatar* condition provides a visual representation of the peer. As stated in H0, this is expected to foster the perception of being with a real person rather than an abstract counterpart. This minimal social visibility introduces non-verbal cues and opportunities for connection, which we expect to strengthen prosocial motivation. Relevant (strategy-related) pre-play face-to-face communication is known to reliably increase cooperation (Isaac & Walker, 1988; Orbell et al., 1988) and coordination (Moreno & Wooders, 1998) with some studies demonstrating that this means of communication is stronger than written forms (Brosig et al., 2003, 2004). At the same time, restricted non-game related communication (Roth et al., 1995; Zultan, 2012)—as well as simple visual identification of the human partner (Bohnet & Frey, 1999)—have also been found to affect cooperative behaviour. We anticipate that participants exposed to avatar representations will demonstrate greater contribution and accuracy in their recycling task, leading to more accurate item recycling compared to the *No Pre-Task Avatar* condition. This improved performance should contribute more effectively to the provision of the public good, ultimately increasing the probability that peers successfully reach the threshold. The simple avatar representation of the peer and his physical materialisation could increase the sense of shared social responsibility. These considerations informed the formulation of the following two experimental hypotheses:

- **H1.a**: The number of correctly recycled items is significantly higher in the *Pre-Task Avatar* treatment than in the *No Pre-Task Avatar* treatment.

- **H1.b**: The probability of reaching the recycling threshold is significantly higher in the *Pre-Task Avatar* treatment than in the *No Pre-Task Avatar* treatment.

---

[8]While Mol (2019) suggests controlling for video game familiarity in VR studies, we did not include this measure as all participants completed an extensive standardized training session designed to ensure uniform preparation levels across participants.

[9]The experimental design and hypotheses were pre-registered at AsPredicted (#226759) prior to data collection. Pre-registration is available at https://aspredicted.org/9h6v-b4hc.pdf

In our study, the questionnaire evaluating individual pro-environmental behavior allowed us to define an environmental score. Individuals who report higher environmental concern should be more motivated to avoid missorting or failing to recycle altogether, thereby acting in accordance with their stated attitudes. Consequently, we expected a positive association between the environmental score and the number of correctly recycled items, regardless of experimental condition. This led us to formulate the following hypothesis:

- **H1.c**: The environmental score is positively correlated with the number of correctly recycled items.

The first dimension of presence we measure, the sense of spatial presence within the virtual environment is assessed through two separate indices: one based on the full IPQ questionnaire composed of 14 items, and the other based on the 10 extracted items from the revised Witmer and Singer questionnaire. Higher reported VR presence on either index reflects a stronger sense of being physically present and a stronger ease of acting in the virtual environment. Consequently, individuals scoring higher should navigate more fluently and handle virtual objects more effectively, thereby yielding more correctly recycled items independently of the condition.

The second dimension of presence we measure which is the sense of co-presence is composed of an 8-item questionnaire, using the first part of the Networked Minds Social Presence Inventory. We supposed that when co-presence is higher, the participant would feel more connected to the peer in the VR environment, and therefore it would reflect on the cooperation score and consequently on the number of correctly recycled items.

- **H1.d**: The sense of presence in VR is positively correlated with the number of correctly recycled items.

Finally, we observe that recycling demands greater real-effort and is more time-consuming than not recycling. As the deadline approaches, limited time, combined with the attentional demands and precision required for correct sorting, reduces anticipated accuracy and our expectation for successfully reaching the threshold. Consequently this could affect our choice to attempt recycling. Taken together, these factors should generate a negative time trend in the probability of choosing the recycling option. Moreover, when participants are in a hurry, the likelihood of making recycling mistakes should increase accordingly. Based on these considerations, we formulated the final two experimental hypotheses:

- **H2.a**: The likelihood of choosing to recycle an item decreases over time.

- **H2.b**: Given the choice to recycle, the likelihood of recycling an item correctly decreases over time.

## 5. Experimental Results

### 5.1. Sample characteristics

The experimental study[10] was conducted in May 2025 in Genoa, Italy, at two university locations (Department of Informatics, Bioengineering, Robotics, and Systems Engineering; Faculty of Law). Participants were recruited through fliers distributed in common areas such as hallways, libraries, and the cafeteria. The fliers included a registration link where participants could sign up for available time slots. Our experiment involved 188 participants across 94 sessions, with each session consisting of a pair of two individuals. In the experiment, no incidents, falls or VR sickness were reported, none withdrew due to discomfort, and all participants successfully completed their session. Exactly half of the sessions were assigned to the *No Pre-Task Avatar* treatment (47) and half to the *Pre-Task Avatar* treatment (47).

Table D.7 in Appendix D.1 presents details of our sample characteristics. Statistical tests revealed only one significant difference between treatments: a higher proportion of participants in the *Pre-Task Avatar* treatment belonged to the youngest age group (18–24). This difference does not pose a concern, given that the *Age* variable has no effect in any of the subsequent analyses. Regarding educational background, the majority of student participants were enrolled in STEM fields, particularly computer science, engineering, and mathematics programs.

### 5.2. Sense of co-presence

The *Co-presence* score, self-reported at the end of the experiment was measured on a 5-point Likert scale. We used a one-sided t-test to compare treatments. We report the full distributions, the mean and the median values per treatment in Figure 4. *Co-presence* is higher in the *Pre-Task Avatar* treatment than in the *No Pre-Task Avatar* treatment (*Pre-Task Avatar*: $M = 3.319$, $SD = 1.036$; *No Pre-Task Avatar*: $M = 1.922$, $SD = 0.848$) and it is statistically significant, $t(186) = -10.122$, $p < 0.001$, $d = -1.476$. For robustness, we also conducted a non-parametric Wilcoxon rank-sum test, which yielded consistent results ($z = -8.290$, $p < 0.001$). This represents a large effect, with participants in the *Pre-Task Avatar* condition reporting co-presence levels above the scale midpoint (3.0), while those in the *No Pre-Task Avatar* condition reported below-average *Co-presence* experiences. This result suggests that avatar representation of the peer before performing the main task, even when brief, has a meaningful impact on users' sense of co-presence in virtual environments, providing strong support for our Hypothesis H0.

---

[10]The experimental study was approved prior to data collection by the ethical committee of Université Côte d'Azur (Decision 2025-056).

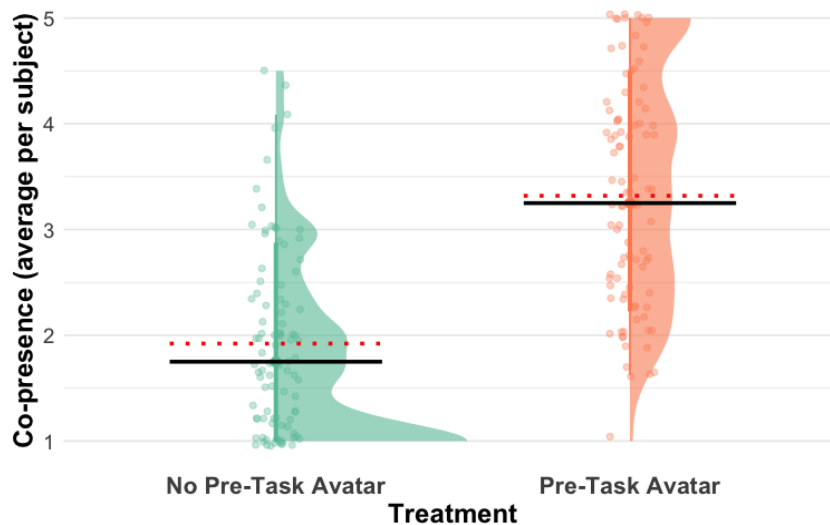

Figure 4: *Co-presence* per subject and by treatment: mean (dotted red line) and median (solid black line)

### 5.3. Number of correctly recycled items and probability of reaching the recycling threshold

For the following analyses, we defined two dependent variables: the number of *Correctly recycled* items per participant, which ranges from 0 to 56, and the binary variable *Recycling threshold*, which equals 1 if the threshold was reached by the pair, and 0 otherwise. Before running the regressions, we conducted a preliminary bivariate analysis. Results are reported in Appendix D.3.

Table 1 reports the OLS regressions predicting the number of *Correctly recycled* items per participant. We can immediately observe that the treatment shows no significant effect in any model, indicating that adding avatar peer representations alone did not increase the contribution level. Thus, our regression analysis does not support Hypothesis H1.a. Similarly, the *Environmental score* (Model 7) does not predict recycling behavior (the number of correctly recycled items) in VR, meaning that our analysis also fails to support Hypothesis H1.c. In contrast, questionnaire measures of the VR experience appear more relevant. *Co-presence* (Model 3) shows a positive and weakly significant effect when entered alone, but this effect disappears once spatial presence measures are included (Models 4–6). The *WS* (Witmer-Singer Presence) index emerges as the most robust predictor of *Correctly recycled*, showing a large and highly significant positive effect across models. Surprisingly, we observe a negative effect of *IPQ* (Model 8), which clearly comes from the subscale measuring experimental realism (*REAL*, Models 5 and 9). Given the strong positive result for *WS*, we conclude that our analysis supports Hypothesis H1.d.

Examining Table 1 reveals several additional noteworthy findings. First, we observe a significant effect of *Gender*, with male participants demonstrating higher recycling accuracy compared to female participants. Second, *Nationality* emerged as a significant predictor, with Italian participants showing superior recycling performance. This latter finding likely reflects Italy's considerable efforts in promoting and enhancing recycling practices (Circular Economy Network & ENEA, 2024). Moreover, it may be attributed to the fact that our experimental design mirrors the recycling system used in Italy (types of bins,

Table 1: OLS regressions: *Correctly recycled* items; Subject-level

|  | (1) Treatment | (2) + Controls | (3) + Co-presence | (4) + IPQ overall | (5) + IPQ sub-categories | (6) + WS | (7) + Env. score | (8) Full (overall IPQ) | (9) Full (IPQ sub-categories) |
|---|---|---|---|---|---|---|---|---|---|
| **Treatment (No Pre-Task Avatar=0; Pre-Task Avatar=1)** | 1.000 (1.414) | -0.142 (1.237) | -1.716 (1.552) | -0.153 (1.238) | -0.071 (1.223) | -0.670 (1.138) | -0.195 (1.242) | -1.536 (1.391) | -1.383 (1.391) |
| **Co-presence** |  |  | 1.124* (0.676) |  |  |  |  | 0.465 (0.624) | 0.529 (0.632) |
| **IPQ overall** |  |  |  | -0.718 (0.897) |  |  |  | -3.509*** (0.886) |  |
| **IPQ: General Presence (GP)** |  |  |  |  | 0.357 (0.730) |  |  |  | -0.052 (0.667) |
| **IPQ: Spatial Presence (SP)** |  |  |  |  | 1.428 (0.906) |  |  |  | -0.658 (0.864) |
| **IPQ: Involvement (INV)** |  |  |  |  | -0.848 (0.600) |  |  |  | -0.613 (0.546) |
| **IPQ: Experienced Realism (REAL)** |  |  |  |  | -2.033** (0.830) |  |  |  | -2.745*** (0.750) |
| **WS** |  |  |  |  |  | 4.198*** (0.710) |  | 5.360*** (0.772) | 5.321*** (0.796) |
| **Environmental score** |  |  |  |  |  |  | 0.844 (1.443) | 0.288 (1.306) | 0.353 (1.311) |
| **Age** |  | -1.315 (0.876) | -1.193 (0.875) | -1.385 (0.881) | -1.407 (0.878) | -0.761 (0.809) | -1.310 (0.878) | -0.896 (0.784) | -0.996 (0.788) |
| **Gender: Male=0; Female=1** |  | -7.605*** (1.275) | -7.374*** (1.276) | -7.767*** (1.292) | -7.045*** (1.292) | -5.518*** (1.221) | -7.681*** (1.284) | -5.659*** (1.193) | -5.425*** (1.193) |
| **Nationality: Italian=0; Other=1** |  | -5.456*** (1.733) | -5.759*** (1.734) | -5.285*** (1.748) | -4.199** (1.766) | -4.619*** (1.596) | -5.172*** (1.803) | -3.576** (1.646) | -2.682 (1.672) |
| **Studies (level)** |  | -0.622 (0.843) | -0.620 (0.839) | -0.513 (0.855) | -0.498 (0.849) | -0.999 (0.776) | -0.668 (0.849) | -0.588 (0.759) | -0.630 (0.761) |
| **VR use frequency** |  | 2.346*** (0.698) | 2.157*** (0.704) | 2.373*** (0.700) | 2.038*** (0.702) | 1.707*** (0.650) | 2.370*** (0.701) | 1.594** (0.636) | 1.514** (0.638) |
| **Risk preference** |  | -0.088 (0.377) | -0.154 (0.378) | -0.080 (0.378) | -0.037 (0.373) | -0.272 (0.348) | -0.081 (0.378) | -0.309 (0.338) | -0.262 (0.337) |
| Num.Obs. | 188 | 188 | 188 | 188 | 188 | 188 | 188 | 188 | 188 |
| $R^2$ | 0.003 | 0.281 | 0.292 | 0.283 | 0.321 | 0.398 | 0.282 | 0.448 | 0.467 |

Notes: Coefficients with standard errors in parentheses. $^*p < 0.10$, $^{**}p < 0.05$, $^{***}p < 0.01$.

waste categories). Even if the color coding was deliberately chosen to be different from the traditional one in order to try to limit the advantages for locals, Italian participants may still have benefited from greater familiarity with the recycling infrastructure presented in the virtual environment. Finally, we found that participants with higher *VR use frequency*—that is, those more accustomed to using VR devices—performed significantly better on the task. This result suggests that prior experience with VR technology enhances users' dexterity and comfort with the interface, thereby improving task performance.

Table 2 reports the results of logit regressions predicting the probability of reaching the recycling threshold. The only stable predictors are *WS*, which shows a significant positive effect and confirms this measure as a robust predictor of both recycling accuracy and threshold achievement, and *Gender*, with male participants consistently performing better on the task. Importantly, as the treatment shows no significant effect, Hypothesis H1.b is also not supported by our experimental analysis.

While the analyses presented above do not support our initial hypotheses regarding treatment effects, the following section presents an alternative analytical approach that captures a different aspect recycling intentions. This analysis reveals a substantial treatment effect that becomes apparent when separating between motivation from task proficiency.

### 5.4. Number of items in the black bin

Although our experimental hypotheses were originally designed around *Correctly recycled* as a measure of recycling contribution, our regression results reveal a more nuanced picture. As discussed in Section 3.3, when presenting our experimental design, our task is a real-effort task, meaning that recycling accuracy depends not only on intentions but also on participants' individual abilities and familiarity with the virtual

Table 2: Logit regressions: Reaching the recycling threshold (0 = No; 1 = Yes); Pair-level

|  | (1) Treatment | (2) + Controls | (3) + Co-presence | (4) + IPQ overall | (5) + IPQ sub-categories | (6) + WS | (7) + Env. score | (8) Full (overall IPQ) | (9) Full (IPQ sub-categories) |
|---|---|---|---|---|---|---|---|---|---|
| **Treatment (No Pre-Task Avatar=0; Pre-Task Avatar=1)** | 1.989 (0.840) | 1.928 (0.885) | 1.262 (0.632) | 1.928 (0.889) | 2.054 (0.994) | 1.844 (0.852) | 1.944 (0.891) | 1.335 (0.666) | 1.368 (0.757) |
| **Co-presence** | | | 1.368* (0.249) | | | | | 1.285 (0.253) | 1.334 (0.292) |
| **IPQ overall** | | | | 1.123 (0.256) | | | | 0.792 (0.226) | |
| **IPQ: General Presence (GP)** | | | | | 0.866 (0.175) | | | | 0.826 (0.173) |
| **IPQ: Spatial Presence (SP)** | | | | | 1.872** (0.489) | | | | 1.451 (0.409) |
| **IPQ: Involvement (INV)** | | | | | 1.004 (0.179) | | | | 1.055 (0.196) |
| **IPQ: Experienced Realism (REAL)** | | | | | 0.639 (0.185) | | | | 0.563* (0.177) |
| **WS** | | | | | | 1.884*** (0.388) | | 1.999*** (0.476) | 1.929** (0.508) |
| **Environmental score** | | | | | | | 0.897 (0.327) | 0.719 (0.298) | 0.765 (0.342) |
| **Age** | | 0.697 (0.157) | 0.719 (0.161) | 0.703 (0.159) | 0.649* (0.160) | 0.735 (0.178) | 0.695 (0.157) | 0.736 (0.173) | 0.674* (0.161) |
| **Gender: Male=0; Female=1** | | 0.189*** (0.069) | 0.197*** (0.072) | 0.193*** (0.071) | 0.209*** (0.083) | 0.237*** (0.091) | 0.190*** (0.071) | 0.242*** (0.093) | 0.247*** (0.105) |
| **Nationality: Italian=0; Other=1** | | 0.387 (0.247) | 0.354 (0.229) | 0.374 (0.247) | 0.510 (0.328) | 0.439 (0.288) | 0.370 (0.243) | 0.382 (0.284) | 0.523 (0.384) |
| **Studies (level)** | | 0.961 (0.227) | 0.955 (0.224) | 0.948 (0.224) | 0.980 (0.248) | 0.922 (0.228) | 0.969 (0.231) | 0.966 (0.235) | 0.973 (0.250) |
| **VR use frequency** | | 1.347* (0.239) | 1.289 (0.237) | 1.340 (0.238) | 1.237 (0.233) | 1.232 (0.235) | 1.343 (0.241) | 1.183 (0.231) | 1.123 (0.228) |
| **Risk preference** | | 1.045 (0.105) | 1.026 (0.105) | 1.043 (0.105) | 1.072 (0.119) | 1.018 (0.106) | 1.044 (0.105) | 1.006 (0.105) | 1.032 (0.118) |
| Num.Obs. | 188 | 188 | 188 | 188 | 188 | 188 | 188 | 188 | 188 |
| McFadden pseudo-$R^2$ | 0.021 | 0.163 | 0.173 | 0.163 | 0.196 | 0.195 | 0.163 | 0.205 | 0.232 |

Notes: Odds ratios reported; *standard errors clustered by Session (HC1)* in parentheses. $^*p < 0.10$, $^{**}p < 0.05$, $^{***}p < 0.01$.

environment. This is clearly reflected in the regression analyses, where factors such as *VR use frequency* and *WS* emerged as strong predictors of performance in both treatment conditions. In particular, we found that participants' comfort with VR technology and their ease in manipulating virtual objects significantly affected task completion, both when an avatar representation of the peer is available or not.

To better isolate the treatment effect on recycling behavior from these confounding ability-related factors, we conducted an analysis focusing on participants' choices to use the black bin. We believe this measure more directly captures contribution intentions, as selecting the black bin for non-recyclable waste requires minimal manual ability (it was the easiest to reach, being positioned in front of the player in the lower row of bins) but reflects a deliberate choice not to contribute to the collective goal. Unlike overall recycling accuracy, which conflates cooperative motivation with task proficiency, black bin usage provides a cleaner indicator of participants' unwillingness to contribute, independent of their skill level in the real-effort task.

Then, we defined the dependent variable *Black bin* per participant, which ranges from 0 to 56. Table 3 reports the OLS regressions predicting this variable per participant.

Table 3: OLS regressions: *Black bin* items; Subject-level

| | (1) Treatment | (2) + Controls | (3) + Co-presence | (4) + IPQ overall | (5) + IPQ sub-categories | (6) + WS | (7) + Env. score | (8) Full (overall IPQ) | (9) Full (IPQ sub-categories) |
|---|---|---|---|---|---|---|---|---|---|
| **Treatment (No Pre-Task Avatar=0; Pre-Task Avatar=1)** | 0.755 | 0.869 | 2.010*** | 0.860 | 0.978* | 0.994* | 0.968* | 1.848*** | 1.928*** |
| | (0.575) | (0.575) | (0.714) | (0.573) | (0.576) | (0.566) | (0.569) | (0.707) | (0.716) |
| **Co-presence** | | | -0.815*** | | | | | -0.576* | -0.579* |
| | | | (0.311) | | | | | (0.317) | (0.325) |
| **IPQ overall** | | | | -0.610 | | | | 0.018 | |
| | | | | (0.415) | | | | (0.451) | |
| **IPQ: General Presence (GP)** | | | | | 0.053 | | | | 0.288 |
| | | | | | (0.344) | | | | (0.343) |
| **IPQ: Spatial Presence (SP)** | | | | | -0.456 | | | | -0.062 |
| | | | | | (0.427) | | | | (0.444) |
| **IPQ: Involvement (INV)** | | | | | 0.210 | | | | 0.057 |
| | | | | | (0.283) | | | | (0.281) |
| **IPQ: Experienced Realism (REAL)** | | | | | -0.652* | | | | -0.468 |
| | | | | | (0.391) | | | | (0.386) |
| **WS** | | | | | | -1.001*** | | -0.772* | -0.765* |
| | | | | | | (0.353) | | (0.393) | (0.410) |
| **Environmental score** | | | | | | | -1.591** | -1.218* | -1.255* |
| | | | | | | | (0.661) | (0.664) | (0.675) |
| **Age** | | 0.328 | 0.239 | 0.268 | 0.236 | 0.195 | 0.318 | 0.157 | 0.176 |
| | | (0.408) | (0.402) | (0.408) | (0.414) | (0.402) | (0.402) | (0.399) | (0.405) |
| **Gender: Male=0; Female=1** | | 0.461 | 0.293 | 0.324 | 0.252 | -0.037 | 0.604 | 0.072 | 0.129 |
| | | (0.593) | (0.587) | (0.598) | (0.609) | (0.608) | (0.588) | (0.607) | (0.614) |
| **Nationality: Italian=0; Other=1** | | 1.706** | 1.926** | 1.852** | 2.047** | 1.507* | 1.170 | 1.292 | 1.503* |
| | | (0.806) | (0.798) | (0.810) | (0.832) | (0.794) | (0.826) | (0.837) | (0.860) |
| **Studies (level)** | | -0.646 | -0.647* | -0.554 | -0.600 | -0.556 | -0.559 | -0.513 | -0.564 |
| | | (0.392) | (0.386) | (0.396) | (0.400) | (0.386) | (0.389) | (0.386) | (0.391) |
| **VR use frequency** | | -0.423 | -0.286 | -0.400 | -0.337 | -0.271 | -0.470 | -0.245 | -0.242 |
| | | (0.325) | (0.324) | (0.324) | (0.330) | (0.323) | (0.321) | (0.323) | (0.328) |
| **Risk preference** | | -0.240 | -0.192 | -0.233 | -0.222 | -0.196 | -0.253 | -0.182 | -0.182 |
| | | (0.176) | (0.174) | (0.175) | (0.176) | (0.173) | (0.173) | (0.172) | (0.173) |
| Num.Obs. | 188 | 188 | 188 | 188 | 188 | 188 | 188 | 188 | 188 |
| $R^2$ | 0.009 | 0.065 | 0.100 | 0.076 | 0.094 | 0.105 | 0.094 | 0.142 | 0.152 |

Notes: Coefficients with standard errors in parentheses. $^{*}p < 0.10$, $^{**}p < 0.05$, $^{***}p < 0.01$.

The regression reveals a significant and unexpected treatment effect: participants in the *Pre-Task Avatar* treatment placed significantly more items in the black bin, indicating lower contributions when having a preliminary brief interaction with their peer's avatar. Although this contradicts our initial hypothesis, we show in the next section that this result reflects improved coordination in the *Pre-Task Avatar* treatment.

### 5.5. Social welfare analysis

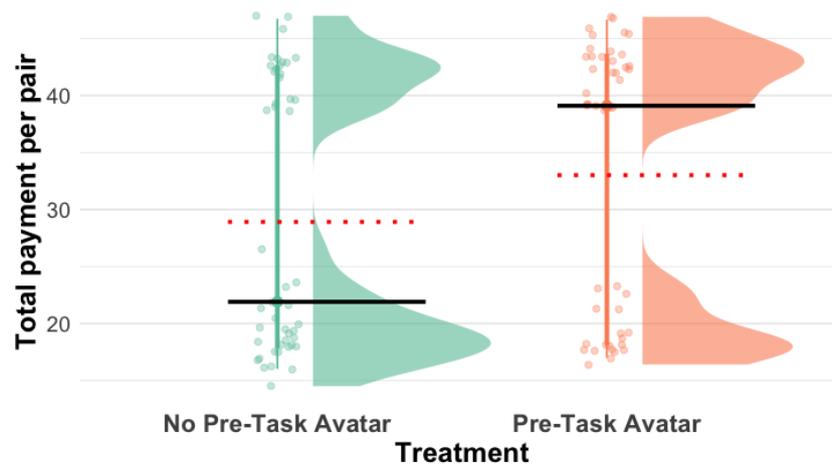

Figure 5: Total payment distribution per pair by treatment: mean (dotted red line) and median (solid black line)

Figure 5 represents the total payment distribution per pair in each of the two conditions. The mean payment (dotted red line) is significantly higher in the *Pre-Task Avatar* treatment ($M = 33.011$, $SD = 11.843$) compared to the *No Pre-Task Avatar* treatment ($M = 28.909$, $SD = 11.819$). The observation was confirmed by a one-sided t-test : $t(92) = -1.681$, $p < 0.05$, $d = -0.346$.

This result should be interpreted within the structure of threshold coordination games. In such settings, contributing beyond the threshold is suboptimal, as wasteful effort reduces individual payoffs without increasing collective benefits. The lower individual contributions observed in the *Pre-Task Avatar* treatment therefore reflect improved strategic coordination rather than reduced cooperation. Pairs in this treatment achieved significantly higher social welfare (defined as the sum of pair payoffs), demonstrating that brief avatar interaction enabled participants to coordinate more efficiently by avoiding over-contribution. This finding reveals a critical distinction: in threshold collective action, effective design features improve coordination quality, not necessarily contribution quantity.

### 5.6. Time pressure analysis

Table 4 presents the results of logistic regressions examining the decision to recycle items (placing them in colored bins, independently of the fact participants do that correctly or not) versus discarding them in the black bin. Here, the dependent variable is *Item recycling bin*, defined per item and per subject, and restricted to those items that have been put in a bin (thus, excluding all items that remained on the desk or on the floor). The findings reveal several important patterns.

Table 4: Logit regressions: Item recycling bin (0 = No, 1 = Yes); Subject-level

|  | (1) Time only | (2) + Controls | (3) + Treatment | (4) + Co-presence | (5) + IPQ overall | (6) + IPQ sub-categories | (7) + WS | (8) + Env. score | (9) Full (overall IPQ) | (10) Full (IPQ sub-categories) |
|---|---|---|---|---|---|---|---|---|---|---|
| **Time (ms since start)** | 0.987*** | 0.987*** | 0.986*** | 0.987*** | 0.987*** | 0.987*** | 0.987*** | 0.987*** | 0.986*** | 0.986*** |
|  | (0.002) | (0.002) | (0.002) | (0.002) | (0.002) | (0.002) | (0.002) | (0.002) | (0.002) | (0.002) |
| **Treatment (No Pre-Task Avatar=0; Pre-Task Avatar=1)** |  |  | 0.433*** |  |  |  |  |  | 0.263*** | 0.265*** |
|  |  |  | (0.135) |  |  |  |  |  | (0.091) | (0.090) |
| **Co-presence** |  |  |  | 1.094 |  |  |  |  | 1.384* | 1.377* |
|  |  |  |  | (0.146) |  |  |  |  | (0.233) | (0.244) |
| **IPQ overall** |  |  |  |  | 1.431* |  |  |  | 0.985 |  |
|  |  |  |  |  | (0.262) |  |  |  | (0.203) |  |
| **IPQ: General Presence (GP)** |  |  |  |  |  | 1.016 |  |  |  | 0.849 |
|  |  |  |  |  |  | (0.181) |  |  |  | (0.162) |
| **IPQ: Spatial Presence (SP)** |  |  |  |  |  | 1.472* |  |  |  | 1.186 |
|  |  |  |  |  |  | (0.301) |  |  |  | (0.271) |
| **IPQ: Involvement (INV)** |  |  |  |  |  | 0.861 |  |  |  | 0.948 |
|  |  |  |  |  |  | (0.154) |  |  |  | (0.165) |
| **IPQ: Experienced Realism (REAL)** |  |  |  |  |  | 1.199 |  |  |  | 1.129 |
|  |  |  |  |  |  | (0.217) |  |  |  | (0.226) |
| **WS** |  |  |  |  |  |  | 1.739*** |  | 1.643*** | 1.564** |
|  |  |  |  |  |  |  | (0.297) |  | (0.295) | (0.294) |
| **Environmental score** |  |  |  |  |  |  |  | 2.305** | 2.193** | 2.270** |
|  |  |  |  |  |  |  |  | (0.827) | (0.784) | (0.749) |
| **Age** |  | 0.946 | 0.856 | 0.975 | 1.008 | 1.026 | 1.070 | 0.968 | 0.993 | 0.977 |
|  |  | (0.203) | (0.173) | (0.219) | (0.223) | (0.235) | (0.243) | (0.222) | (0.217) | (0.214) |
| **Gender: Male=0; Female=1** |  | 0.807 | 0.781 | 0.815 | 0.878 | 0.961 | 1.122 | 0.730 | 0.955 | 0.911 |
|  |  | (0.259) | (0.244) | (0.267) | (0.293) | (0.316) | (0.420) | (0.230) | (0.323) | (0.322) |
| **Nationality: Italian=0; Other=1** |  | 0.449* | 0.381** | 0.451* | 0.418** | 0.401** | 0.526 | 0.531 | 0.491 | 0.493 |
|  |  | (0.200) | (0.172) | (0.199) | (0.184) | (0.186) | (0.229) | (0.268) | (0.266) | (0.268) |
| **Studies (level)** |  | 1.472* | 1.562** | 1.452* | 1.367 | 1.382 | 1.305 | 1.384 | 1.293 | 1.331 |
|  |  | (0.318) | (0.345) | (0.308) | (0.288) | (0.306) | (0.285) | (0.342) | (0.305) | (0.309) |
| **VR use frequency** |  | 1.350* | 1.398* | 1.321 | 1.346* | 1.295 | 1.265 | 1.387* | 1.222 | 1.213 |
|  |  | (0.223) | (0.244) | (0.231) | (0.225) | (0.225) | (0.217) | (0.241) | (0.234) | (0.224) |
| **Risk preference** |  | 1.123 | 1.124 | 1.118 | 1.114 | 1.136 | 1.113 | 1.137 | 1.107 | 1.115 |
|  |  | (0.111) | (0.112) | (0.111) | (0.109) | (0.113) | (0.109) | (0.118) | (0.122) | (0.124) |
| Num.Obs. | 8962 | 8962 | 8962 | 8962 | 8962 | 8962 | 8962 | 8962 | 8962 | 8962 |
| McFadden pseudo-$R^2$ | 0.115 | 0.148 | 0.164 | 0.149 | 0.154 | 0.164 | 0.169 | 0.163 | 0.208 | 0.211 |

Notes: Odds ratios reported; *standard errors clustered by participant (HC1)* in parentheses. $^*p < 0.10$, $^{**}p < 0.05$, $^{***}p < 0.01$.

First, we observe a strong and significant treatment effect. Participants in the *Pre-Task Avatar* treatment are significantly less likely to recycle items compared to those in the *No Pre-Task Avatar* treatment. This result aligns with our previous findings and should be interpreted within the coordination framework of the threshold public good game. As discussed earlier, contributing beyond the threshold is suboptimal, and the lower recycling rate in the *Pre-Task Avatar* treatment reflects better coordination rather than reduced environmental engagement. This interpretation is further supported by the significantly higher social welfare observed in this treatment (as shown in Section 5.5).

Second, the temporal pattern is highly significant across all specifications, where the odds of recycling decrease over time, thus supporting Hypothesis H2.a. This suggests that participants become more strategic as the experiment progresses, possibly learning to avoid over-contribution. Also, time pressure certainly plays a role, with participants getting more and more engaged in completing the task and cleaning the desk. The consistency of this effect across all models indicates its robustness.

Third, individual characteristics play important roles. *WS* again positively predicts recycling behavior. Similarly, *Environmental score* strongly predicts recycling decisions: participants with higher environmental scores are more than twice as likely to recycle items.

Again, *Co-presence* is marginally significant in the full models, suggesting that feeling the presence of another participant increases the propensity to recycle, though this effect is weaker than the direct treatment effect. Among socio-demographic variables, *Nationality* shows a consistent negative effect (non-Italian participants are less likely to recycle), while *VR use frequency* and *Studies* positively correlate with recycling behavior, though these effects become non-significant in the full specifications.

Table 5: Logit regressions: Item correctly recycled(0 = Incorrect bin; 1 = Correct bin); Subject-level

| | (1) Time only | (2) + Controls | (3) + Treatment | (4) + Co-presence | (5) + IPQ overall | (6) + IPQ sub-categories | (7) + WS | (8) + Env. score | (9) Full (overall IPQ) | (10) Full (IPQ sub-categories) |
|---|---|---|---|---|---|---|---|---|---|---|
| **Time (ms since start)** | 1.001 | 1.001 | 1.001 | 1.001 | 1.001 | 1.001 | 1.001 | 1.001 | 1.001 | 1.001 |
| | (0.001) | (0.001) | (0.001) | (0.001) | (0.001) | (0.001) | (0.001) | (0.001) | (0.001) | (0.001) |
| **Treatment (No Pre-Task Avatar=0; Pre-Task Avatar=1)** | | | 0.970 | | | | | | 0.999 | 0.981 |
| | | | (0.153) | | | | | | (0.172) | (0.162) |
| **Co-presence** | | | | 0.981 | | | | | 0.963 | 0.987 |
| | | | | (0.064) | | | | | (0.068) | (0.076) |
| **IPQ overall** | | | | | 0.933 | | | | 0.833 | |
| | | | | | (0.089) | | | | (0.103) | |
| **IPQ: General Presence (GP)** | | | | | | 1.137 | | | | 1.134 |
| | | | | | | (0.117) | | | | (0.117) |
| **IPQ: Spatial Presence (SP)** | | | | | | 0.905 | | | | 0.797* |
| | | | | | | (0.119) | | | | (0.104) |
| **IPQ: Involvement (INV)** | | | | | | 1.012 | | | | 1.019 |
| | | | | | | (0.066) | | | | (0.070) |
| **IPQ: Experienced Realism (REAL)** | | | | | | 0.826* | | | | 0.794** |
| | | | | | | (0.087) | | | | (0.079) |
| **WS** | | | | | | | 1.190* | | 1.298** | 1.372** |
| | | | | | | | (0.120) | | (0.152) | (0.170) |
| **Environmental score** | | | | | | | | 0.936 | 0.925 | 0.930 |
| | | | | | | | | (0.157) | (0.147) | (0.143) |
| **Age** | | 0.879 | 0.877 | 0.876 | 0.870 | 0.885 | 0.907 | 0.878 | 0.885 | 0.907 |
| | | (0.084) | (0.085) | (0.085) | (0.085) | (0.085) | (0.089) | (0.084) | (0.088) | (0.087) |
| **Gender: Male=0; Female=1** | | 0.847 | 0.846 | 0.844 | 0.836 | 0.847 | 0.901 | 0.854 | 0.896 | 0.909 |
| | | (0.121) | (0.122) | (0.122) | (0.118) | (0.122) | (0.137) | (0.124) | (0.140) | (0.138) |
| **Nationality: Italian=0; Other=1** | | 0.384*** | 0.382*** | 0.384*** | 0.391*** | 0.423*** | 0.397*** | 0.375*** | 0.414*** | 0.451*** |
| | | (0.067) | (0.069) | (0.067) | (0.068) | (0.082) | (0.070) | (0.066) | (0.079) | (0.091) |
| **Studies (level)** | | 0.842* | 0.842* | 0.842* | 0.851 | 0.834* | 0.818* | 0.846* | 0.842 | 0.820* |
| | | (0.084) | (0.084) | (0.084) | (0.086) | (0.084) | (0.085) | (0.084) | (0.089) | (0.084) |
| **VR use frequency** | | 1.253*** | 1.252*** | 1.255*** | 1.258*** | 1.271*** | 1.226** | 1.247** | 1.223** | 1.239** |
| | | (0.107) | (0.107) | (0.108) | (0.109) | (0.104) | (0.108) | (0.109) | (0.113) | (0.106) |
| **Risk preference** | | 0.995 | 0.996 | 0.997 | 0.996 | 0.994 | 0.988 | 0.995 | 0.990 | 0.982 |
| | | (0.041) | (0.041) | (0.042) | (0.041) | (0.041) | (0.040) | (0.042) | (0.041) | (0.042) | (0.040) |
| Num.Obs. | 8625 | 8625 | 8625 | 8625 | 8625 | 8625 | 8625 | 8625 | 8625 | 8625 |
| McFadden pseudo-$R^2$ | 0.000 | 0.038 | 0.038 | 0.038 | 0.038 | 0.042 | 0.040 | 0.038 | 0.043 | 0.047 |

Notes: Odds ratios reported; *standard errors clustered by participant (HC1)* in parentheses. $^*p < 0.10$, $^{**}p < 0.05$, $^{***}p < 0.01$.

Table 5 examines the quality of recycling decisions among participants who chose to recycle a given item (i.e., conditional on not using the black bin and on not letting the item unplaced). The dependent variable is *Item correctly recycled*, defined per item and per subject. This analysis reveals distinct patterns from those observed for the recycling decision itself. Most notably, the treatment effect disappears when examining recycling accuracy: there is no significant difference between the *Pre-Task Avatar* and *No Pre-Task Avatar* treatments in terms of correct bin placement. This finding is meaningful–it indicates that while participants in the *Pre-Task Avatar* treatment recycle less frequently (as shown before), when they do recycle, they are equally accurate as those in the *No Pre-Task Avatar* treatment. This dissociation between quantity and quality of recycling further supports our coordination interpretation: participants in the *Pre-Task Avatar* treatment are not less capable or less motivated to recycle correctly; they simply recycle less because they coordinate more efficiently to avoid over-contribution.

The temporal pattern also differs: time since the start of the experiment does not significantly affect recycling accuracy (coefficient ≈ 1.001 across all models), thus in contrast with Hypothesis H2.b. This suggests that while participants learn to recycle less over time, those who continue to recycle maintain consistent accuracy throughout the experiment.

Several variables that predicted recycling behavior show different patterns here. *Environmental score*,

which strongly predicted the decision to recycle, does not significantly affect recycling accuracy. This suggests that environmental attitudes influence whether participants contribute to the public good, but not necessarily their technical ability to do so correctly. Conversely, *WS* remains significant, indicating that prosocial motivation is associated with both more frequent and more accurate recycling.

The IPQ sub-categories reveal interesting patterns. While *Spatial Presence* and *Experienced Realism* negatively predict correct recycling, these effects are modest. One possible interpretation is that participants who felt more spatially present or perceived the virtual environment as more realistic may have been more focused on the social interaction or the virtual experience itself, potentially diverting attention from the technical task of identifying correct bins. Finally, *Nationality* still emerges as a very strong predictor: non-Italian participants are substantially less likely to correctly sort items, even when they choose to recycle. This likely reflects familiarity with Italian recycling rules and sorting systems. *VR use frequency* positively predicts accuracy, suggesting that comfort with the virtual environment facilitates correct task execution.

## 6. Discussion and Conclusion

In this paper, we examined how social representation features in collaborative virtual systems influence coordination in threshold public goods provision. Through a controlled experiment, we investigated whether brief exposure to peer avatar representation increases perceived co-presence and improves coordination outcomes in a real-effort recycling task. Participants in the *Pre-Task Avatar* treatment briefly met and interacted through avatar representations in a shared virtual space before the main task that is carried out alone, while those in the *No Pre-Task Avatar* treatment completed an equivalent preliminary activity individually without peer visibility. Virtual Reality enabled us to reintroduce realistic social context into laboratory experimentation while maintaining experimental control, isolating the effect of co-presence by excluding verbal communication during the task.

Our findings reveal important nuances. Brief avatar exposure significantly increases perceived co-presence, demonstrating that minimal design interventions can effectively create social presence. However, avatar representation did not directly increase individual contribution quantity. Instead, participants in the *Pre-Task Avatar* treatment placed significantly more items in the black bin–reducing recycling effort–yet pairs achieved significantly higher social welfare. This pattern indicates improved strategic coordination through reduced over-contribution beyond the threshold. Additionally, VR presence–users' sense of "being there" in the virtual environment–emerged as a strong predictor of task performance, suggesting that immersion quality may matter as much as social features for collaborative tasks.

This research makes several contributions to IS theory, methodology, and practice. We provide actionable insights for designers of collaborative platforms supporting crowdfunding, sustainability initiatives, community organizing, and other collective action domains. For organizations deploying collaborative platforms, understanding which design features facilitate coordination can maximize platform effectiveness while managing development costs. We show that minimal social exposure through avatar representation can affect strategic coordination even without sustained communication, challenging assumptions that rich, ongoing interaction is necessary for presence to influence behavior. This finding has important implications: platforms do not need to invest in sophisticated communication systems or sustained interaction capabilities to facilitate coordination. Brief visual co-presence may suffice for certain collective action contexts. Importantly, our findings reveal that platform designers should also prioritize investments in immersion quality and interface usability and not only on elaborate social features when supporting task-oriented collaboration. Finally, our findings inform the debate about anonymity versus identification in collaborative platforms, demonstrating that avatar representations can provide social cues sufficient for coordination without revealing actual identity.

In addition, we also establish Virtual Reality as a rigorous experimental methodology for IS research on collaborative systems. VR enables unprecedented control over system features while maintaining ecological validity through immersive, behaviorally-consequential tasks. Our approach demonstrates how researchers can isolate design effects that would be confounded in field studies of real platforms while preserving the richness of collaborative interaction. By implementing a real-effort task requiring both strategic decision-making and physical execution under time pressure, we provide a methodological template for future IS research examining user behavior in demanding collaborative contexts. This approach addresses a persistent challenge in IS research: studying how system features influence behavior in realistic settings without sacrificing experimental control. The rich behavioral data VR enables—including movement patterns, temporal dynamics, and task execution quality—offers insights impossible to capture in traditional laboratory or field studies.

We conclude by observing that, by implementing a threshold public goods game, our study demonstrates how VR can be applied as a rigorous experimental tool also within the field of experimental economics. So far, economists have relied mainly on two experimental traditions: laboratory experiments, ensuring strict control and internal validity, and field experiments that emphasise ecological validity and the observation of behaviour in natural contexts. VR extends the possibilities of conventional laboratory and field settings by combining high experimental control with realistic environments (Fiore et al., 2009; Harrison et al., 2011, 2020; Innocenti, 2017; Mol, 2019), a methodological approach that Innocenti (2017) describes as "framed field experiments."

Unlike standard laboratory methods that rely on abstract or hypothetical scenarios, VR enables to design immersive, context-rich environments in which participants make decisions and perform actions that closely resemble situations they are facing in real-life. This approach reduces participants' reliance on mental reconstruction and facilitates a more natural engagement with experimental tasks (Mol, 2019). As highlighted by Sutan et al. (2022), Attanasi et al. (2025), Harrison et al. (2020), Mol et al. (2022), immersive technologies have the ability to provide vivid visualizations, enhancing participants' understanding of complex socio-economic issues—including environmental challenges such as waste accumulation, plastic pollution, wildfires, and climate-related disasters, but also the destructive consequences of war (Gürerk & Kasulke, 2021). Moreover, VR enables participants to perform real-effort tasks that require authentic physical engagement with virtual objects (Graff et al., 2021; Gürerk et al., 2019), allowing participants to experience the tangible consequences of their decisions as they affect the virtual environment, producing outcomes that extend beyond simple monetary incentives.

VR introduces also a novel dimension by enabling participants to experience embodiment and interact with computer-controlled agents or user-controlled avatars. Researchers in economics have explored how the mere presence of a virtual agent can affect cheating behavior (Mol et al., 2020), but also how changing the productivity of a virtual peer can affect performance behavior in dynamic tournaments (Graff et al., 2021; Gürerk et al., 2019).

This added layer of social realism also raises important methodological considerations about anonymity, which is a fundamental principle in laboratory experiments (Charness & Gneezy, 2008). However, complete anonymity is unnatural in everyday social interactions, leading to a weakening of interpersonal ties, as identifiable characteristics of others are absent. In increasingly anonymous settings, participants tend toward stronger self-interested behavior, aligned with payoff maximization (Franzen & Pointner, 2012). Conversely, reducing social distance through increased visibility leads to higher levels of other-regarding behavior (Haruvy et al., 2017), potentially through mechanisms such as fear of retaliation, guilt aversion, adherence to social norms, or reputation concerns. Empirical evidence further shows that revealing identities—through family names (Charness & Gneezy, 2008) or photographs (Andreoni & Petrie, 2004)—influences cooperation and generosity, although such identification risks introducing uncontrolled variation in personal characteristics (e.g., age, gender, ethnicity, attractiveness) that can affect prosocial behavior through ingroup–outgroup dynamics or stereotyping effects (Fershtman & Gneezy, 2001). While conventional experiments typically ensure anonymity to prevent social influence, VR offers a middle ground: digital representations allow researchers to reintroduce controlled social cues—such as avatar appearance or body movement—providing a sense of social presence without revealing participants' actual identities.

Future research could examine how variations in avatar features—such as gender, age, or ethnicity (van der Land et al., 2015), or how more closely an avatar resembles its user (Suh et al., 2011)—and in the degree of behavioral realism influence cooperation. Another avenue lies in comparing real-effort and stated-choice designs to better understand how tangible effort affects prosocial behavior. Moreover, VR provides new opportunities to study communication under controlled anonymity: while our design excluded verbal interaction to isolate co-presence, future studies could use avatars to offer novel ways to investigate how combined controlled verbal and non-verbal interaction shapes cooperation.

## Acknowledgments

The authors would like to thank Arno Riedl and Paolo Zeppini for helpful comments and suggestions on earlier drafts of this paper. The project has received funding from the Grant Agreement number: 101058636 - ABSolEU - HORIZON-CL4-2021.

# APPENDIX

## A. Questionnaires

### A.1. Environmental score questionnaire

The questionnaire is adapted from Czajkowski et al. (2017). It examines recycling behavior through a model incorporating social pressure, moral motivation, and perceived private costs or effort. For the purposes of this study, items wording was modified to align with our research context, and the final item "Sorting waste will allow to reduce my bills" was replaced with "I recycle my waste" to better reflect our objectives. The 10 items included in the questionnaire are:

Table A.1: List of items for *Environmental score* assessment

| Item | Question | Anchors |
|---|---|---|
| 1 | Recycling is better for the environment than throwing everything into a general waste bin. | |
| 2 | Recycling requires too much effort compared to the benefits it provides. | |
| 3 | Personally, recycling would give me satisfaction. | 1 = I definitely disagree |
| 4 | When I recycle, I do it carefully. | 5 = I definitely agree |
| 5 | I know how to properly sort my waste for recycling. | |
| 6 | Recycling is a moral/ethical duty for me. | |
| 7 | I think people would judge me negatively if I didn't recycle. | |
| 8 | I judge people negatively who don't recycle. | |
| 9 | It is important for people to make an effort to recycle. | |
| 10 | I recycle my waste. | |

Items were measured on a 5-point Likert scale ranging from "I definitely disagree" to "I definitely agree". The second item has a reverse wording. Its score was calculated by subtracting the answer score from 6. In the original paper, responses are not aggregated into a composite score. For the purpose of our study, we defined an *Environmental score* which represents an average of all 10 items (from 1 to 5).

## A.2. Igroup Presence Questionnaire

The Igroup Presence Questionnaire (IPQ) (Schubert et al., 2001) is a standardized scale for measuring the sense of presence in a virtual environment. It is composed of 14 items divided into four groups: *G = General Presence* (general "sense of being there"), *SP = Spatial Presence* (the sense of being physically present in the virtual environment), *INV = Involvement* (measuring the attention devoted to the virtual environment and the involvement experienced), and *REAL = Experienced Realism* (measuring the subjective experience of realism in the virtual environment). The questions were ordered according to the official form of the questionnaire, available at: https://www.igroup.org/pq/ipq/ipq_english.htm. The 14 items included in the questionnaire are:

Table A.2: List of items for *IPQ* assessment

| Item | Question | Anchors |
|---|---|---|
| G | In the computer generated world I had a sense of "being there" | 1 = not at all<br>7 = very much |
| SP1 | Somehow I felt that the virtual world surrounded me. | 1 = fully disagree<br>7 = fully agree |
| SP2 | I felt like I was just perceiving pictures | |
| SP3 | I did not feel present in the virtual space | 1 = did not feel present<br>7 = felt present |
| SP4 | I had a sense of acting in the virtual space, rather than operating something from outside. | 1 = fully disagree<br>7 = fully agree |
| SP5 | I felt present in the virtual space. | |
| INV1 | How aware were you of the real world surrounding while navigating in the virtual world? | 1 = extremely aware<br>7 = not aware at all |
| INV2 | I was not aware of my real environment. | 1 = fully disagree<br>7 = fully agree |
| INV3 | I still paid attention to the real environment. | |
| INV4 | I was completely captivated by the virtual world. | |
| REAL1 | How real did the virtual world seem to you? | 1 = completely real<br>7 = not real at all |
| REAL2 | How much did your experience in the virtual environment seem consistent with your real world experience? | 1 = not consistent<br>7 = very consistent |
| REAL3 | How real did the virtual world seem to you? | 1 = imagined world<br>7 = indistinguishable from real world |
| REAL4 | The virtual world seemed more realistic than the real world. | 1 = fully disagree<br>7 = fully agree |

In this study, a 7-point Likert scale ranging from 1 to 7 was used instead of the original -3 to +3 scale, to align with the format of the other questionnaires in the experiment and to reduce potential participant confusion. Three items (SP2, INV3, and REAL1) have a reverse wording. For these items, scores were calculated by subtracting the original response from 8. We defined the average score for each category, *G, SP, INV, REAL* (from 1 to 7), and the *IPQ* index, computed as an average score based on the answers to all 14 items (from 1 to 7).

## A.3. Witmer & Singer Questionnaire

The table below presents the items used in this study based on the WS questionnaire (Witmer & Singer, 1998). The original WS questionnaire includes a total of 32 items. In this study, we used the modified version from the UQO Cyberpsychology Lab, which consists of 19 items (for VEs without sounds/haptic), or 24 items (for VEs with sounds/touch). The used questionnaire consists of 5 main sub-categories: *Realism, Possibility to act, Quality of interface, Possibility to examine, Self-evaluation of performance* and two additional ones: *Sounds, Haptic*. In total, we selected 10 items, two for each of the main sub-categories (*Realism*: items 5 and 10; *Possibility to act*: items 2 and 9; *Quality of interface*: items 17 and 18; *Possibility to examine*: items 12 and 19; *Self-evaluation of performance*: items 15 and 16). The full list of items can be consulted at the following link: https://marketinginvolvement.wordpress.com/wp-content/uploads/2013/12/pq-presence-questionnaire.pdf. The 10 items included in the questionnaire are:

Table A.3: List of items for *WS* assessment

| Item | Question | Anchors |
|---|---|---|
| Item 2 | How responsive was the environment to actions that you initiated? | 1 = not responsive, 7 = completely responsive |
| Item 5 | How natural was the mechanism which controlled movement through the environment? | 1 = extremely artificial, 7 = extremely natural |
| Item 9 | How completely were you able to actively survey or search the environment using vision? | 1 = not at all, 7 = completely |
| Item 10 | How compelling was your sense of moving around inside the virtual environment? | 1 = not compelling, 7 = completely compelling |
| Item 12 | How well could you examine objects from multiple viewpoints? | 1 = not at all, 7 = extremely well |
| Item 15 | How quickly did you adjust to the virtual environment experience? | 1 = not at all, 7 = very quickly |
| Item 16 | How proficient in moving and interacting with the virtual environment did you feel at the end of the experience? | 1 = not proficient, 7 = extremely proficient |
| Item 17 | How much did the visual display quality interfere or distract you from performing assigned tasks or required activities? | 1 = not at all, 7 = very much |
| Item 18 | How much did the control devices interfere with the performance of assigned tasks or required activities? | |
| Item 19 | How well could you concentrate on the assigned tasks or required activities rather than on the mechanisms used to perform those tasks or activities? | 1 = not at all, 7 = completely |

Responses were measured on a 7-point Likert scale, as in the cited version of the questionnaire. Items 17 and 18 have a reverse wording. For these items, scores were calculated by subtracting the response from 8. We then computed the *WS* score as the average of the answers to all 10 items.

## A.4. Networked Minds Social Presence Inventory

The Networked Minds Social Presence Inventory (Biocca & Harms, 2003) is a standard questionnaire used to measure social presence in VR. It is composed of three sections:

1. First order social presence: co-presence, representing "the degree to which the users feel as if they are together in the same space"

2. Second order social presence: psycho-behavioral interaction, "measuring the user perception of attention, emotional contagion, and mutual understanding with their partner or participant".

3. Third order social presence: subjective and Intersubjective Symmetry, "derived from the scales used for first order and second order social presence".

For this study we concentrated on first order social presence: co-presence, which is composed of 8 items divided into two sub-categories: *Perception of self* (Items 1, 3, 5 and 7) and *Perception of the other* (Items 2, 4, 6, 8). The full questionnaire can be consulted at: https://web-archive.southampton.ac.uk/cogprints.org/6742/1/2002_netminds_scales.pdf. The 10 items in the questionnaire are:

Table A.4: List of items for *Co-presence* assessment

| Item | Question | Anchors |
|---|---|---|
| 1 | I often felt that the other participant and I were in the same room together. | |
| 2 | I think the other participant often felt that we were in the same room together. | |
| 3 | I was often aware of the other participant in the room. | |
| 4 | The other participant was often aware of me in the room. | 1 = strongly disagree |
| 5 | I hardly noticed the other participant in the room. | 5 = strongly agree |
| 6 | The other participant hardly noticed me in the room. | |
| 7 | I often felt as if the other participant and I were in different places rather than together in the same room. | |
| 8 | I think the other participant often felt as if we were in different places rather than together in the same room. | |

Responses were measured on a 5-point Likert scale. The 1.2 version of the questionnaire does not specify the response format, however, a previous version 1.0 of the Networked Minds Social Presence Inventory (Biocca & Harms, 2003) used a 7-point Likert scale. In the present study, we focused extensively on first-order social presence between the two treatments *No Pre-Task Avatar* and *Pre-Task Avatar*–co-presence–and we do not conduct a cross-study comparison.

Items 5, 6, 7 and 8 have a reverse wording. For these items, scores were calculated by subtracting the original response from 6. We defined the two variables *Perception self* and *Perception other* as the average to answers to questions in the corresponding categories, and we computed the *Co-presence* score as the average of the answers to all 8 items.

## A.5. Recycling motivation questionnaire

To investigate the underlying motivation for recycling within the context of the virtual task, a set of items was specifically developed for the purposes of this study. The complete set of 7 items employed in the analysis is presented below.

Table A.5: List of items for *Motivation* assessment

| Item | Question | Anchors |
|---|---|---|
| 1 | *(Quantity)* I tried to recycle as much as possible. | 1 = strongly disagree<br>5 = strongly agree |
| 2 | *(Quality)* I tried to recycle in a meticulous way. | |
| 3 | *(Collective good)* I recycled because the collective good is important to me. | |
| 4 | (Private good) I recycled to try to maximize my own gain. | |
| 5 | (Reciprocity) I recycled because I expected my partner to do it. | |
| 6 | (Competition) I recycled out of a sense of competition with my partner. | |
| 7 | (Cooperation) I recycled out of a sense of collaboration with my partner. | |

Each statement was evaluated using a 5-point Likert scale, ranging from 1 (strongly disagree) to 5 (strongly agree). We did not compute an average score based on all items, but instead each item was studied individually.

## A.6. Additional control variables

To investigate for additional control variables, namely, risk preferences and VR use, we asked the following two questions.

Table A.6: Control measures for risk and VR use

| Question | Anchors |
|---|---|
| In general, in life, how willing are you to take risks? (On a scale from 0 to 10) | 0 = Not at all willing to take risks<br>10 = Completely willing to take risks |
| How often do you use Virtual Reality technology? | 1 = I never use Virtual Reality technology<br>2 = I rarely use Virtual Reality technology<br>3 = I sometimes use Virtual Reality technology<br>4 = I often use Virtual Reality technology |

The first statement was evaluated using an 11-point Likert scale, ranging from 0 (not at all willing to take risks) to 10 (completly willing to take risks). The second statement was evaluated on a 4-point Likert scale, ranging from 1 (I never use Virtual Reality technology) to 4 (I often use Virtual Reality technology). This allowed us defining the following two variables: *Risk preferences* and *VR use frequency*.

# B. Instructions

## B.1. Instructions before the experiment

(N.B. Identical instructions in Italian were provided for Italian speakers.)

Welcome to this experiment organized by DIBRIS (University of Genoa) and Université Côte d'Azur.

In this session, you will take part in a decision-making experience in virtual reality, during which you will have the opportunity to earn money based on your choices and those of another participant. Regardless of your decisions or outcomes, you will receive a fixed participation fee of €5. The experiment will last approximately 45 minutes.

You will be immersed in a virtual reality environment and asked to make decisions by interacting with that environment. To help you become familiar with the technology, the session will begin with a brief training phase. Following the training, you will proceed with the timed main experimental task. You and another participant will perform the same task simultaneously, but you will not be able to see or communicate with each other during the task. Full instructions explaining what is expected of you will be provided before the main task begins.

There is no right or wrong strategy in this experiment. You are free to make your own choices.

During the main task, you will be left alone in the room. Once it is over, as soon as instructed by the application, you may remove the virtual reality headset and notify the experimenters. You will then proceed to complete a final questionnaire. At the end, you will receive a cash payment based on your participation and the results obtained.

All decisions and responses will remain anonymous and will be used exclusively for academic research purposes, including conference presentations and scientific publications.

We kindly ask that you switch off your mobile phone for the entire duration of the experience. If you have any questions or encounter any technical issues, please speak to the experimenter. Remember you are free to withdraw from the experiment at any time and without any penalty.

If you are ready, please proceed by putting on the virtual reality equipment.

## B.2. Instructions before the main recycling task

(N.B. Identical instructions in Italian were provided for Italian speakers.)

Here we are at home. Look at this mess—we really need to do some cleaning.

On the desk in front of you and on the floor there are twenty-four objects. Some of these objects can be separated into parts. Different parts may be made of different materials. In total, there are fifty-six parts in front of you.

The materials that the various parts can be made of are: organic, e-waste, plastic, metal, paper, and glass. The material each part is made of is indicated by a label showing the corresponding symbol. If you can't find it immediately, try rotating the object.

This is the list of materials and their symbols. You'll always find it hanging on the wall to your right in case you need it.

During the experiment, the waste items around the desk must be cleaned up, separated into parts when necessary, and disposed of properly. The total time available is five minutes. In another room, your partner will have the exact same task to perform.

For each part, you can choose to throw it in the black bin for unsorted waste or recycle it. Throwing a part into the black bin for unsorted waste will earn you twenty euro cents, while recycling it in the correctly colored bin for its material will earn you ten euro cents. But be careful—if you place a part in the wrong recycling bin, you won't earn anything.

There are also two bonuses you can earn. If at the end of the experiment there are no more objects around the desk, you will receive a three-euro bonus. In addition, if by the end of the experiment you and your partner in the other room have correctly recycled at least seventy-five percent of all parts—that is, eighty-four out of one hundred twelve parts—both of you will receive a ten-euro bonus.

Happy desk cleaning! When you're ready to begin, press the Start button.

## C. Software Specifications

The following provides a detailed overview of the software packages components used to develop and run the VR environment. The Unity VR Multiplayer Project Template was used to streamline the development and to integrate Unity Cloud Multiplayer Services.

- **Unity version**: 2022.3.46f1
- **OpenXR package**: v1.12.0
- **XR Interaction Toolkit package**: v3.0.5
- **Netcode for GameObjects package**: v1.11.0
- **Lobby package**: v1.2.2

## D. Additional results

### D.1. Sample characteristics

### D.2. Additional analyses Section 5.2, role of *Co-presence*

In this section, we report additional analyses about the measure of *Co-presence* in our study. Among the two categories of *Co-presence* (see Section A.4), *Perception of Self* is marginally correlated with recycling performance (*Correctly recycled* variable as defined at the begining of Section 5.3, Spearman's $\rho = 0.141$, $p = 0.054$), as is *Perception of the other* (Spearman's $\rho = 0.156$, $p = 0.032$). A Spearman correlation test further shows a marginally significant correlation between the average *Co-presence* score and *Correctly recycled* items (Spearman's $\rho = 0.157$, $p = 0.031$), showing that individuals who report very high levels of *Co-presence* recycle more objects on average compared to those with lower *Co-presence* scores (see Figure D.1).

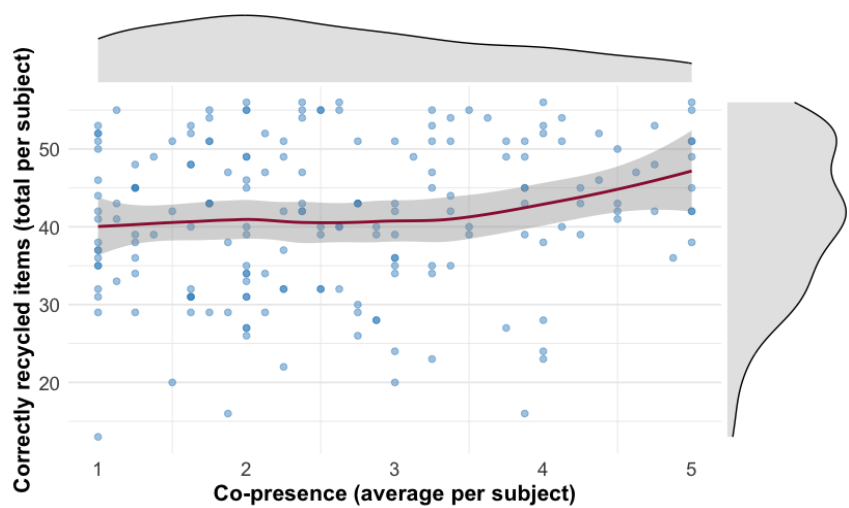

Figure D.1: Total of *Correctly recycled* items (total per subject) and *Co-presence* (per subject)

Although in the *No Pre-Task Avatar* treatment we cannot find the correlation to be marginally significant (Spearman's $\rho = 0.119$, $p = 0.252$), the correlation is significant in the treatment *Pre-Task Avatar* (Spearman's $\rho = 0.201$, $p = 0.052$, see Figure D.2).

Table D.7: Sample characteristics by treatment

|  | **No Pre-Task Avatar** | **Pre-Task Avatar** | **Both** |
|---|---|---|---|
| N (%) | 94 (50.0%) | 94 (50.0%) | 188 (100.0%) |
| *Gender* | | | |
|    Male | 56 (59.6%) | 59 (62.8%) | 115 (61.2%) |
|    Female | 38 (40.4%) | 35 (37.2%) | 73 (38.8%) |
| *Age* | | | |
|    18–24 | 41 (43.6%) | 58 (61.7%) | 99 (52.7%) |
|    25–34 | 44 (46.8%) | 28 (29.8%) | 72 (38.3%) |
|    35–44 | 5 (5.3%) | 6 (6.4%) | 11 (5.9%) |
|    45–54 | 3 (3.2%) | 3 (3.2%) | 6 (3.2%) |
|    55–64 | 1 (1.1%) | 2 (2.1%) | 3 (1.6%) |
| *Nationality* | | | |
|    Italy | 73 (77.7%) | 81 (86.2%) | 154 (81.9%) |
|    Other | 21 (22.3%) | 13 (13.8%) | 34 (18.1%) |
| *Language* | | | |
|    Italian | 76 (80.9%) | 83 (88.3%) | 159 (84.6%) |
|    English | 18 (19.1%) | 11 (11.7%) | 29 (15.4%) |
| *Level of completed studies* | | | |
|    Middle school | 1 (1.1%) | 0 (0.0%) | 1 (0.5%) |
|    High school | 29 (30.9%) | 30 (31.9%) | 59 (31.4%) |
|    Bachelor's | 33 (35.1%) | 39 (41.5%) | 72 (38.3%) |
|    Master's | 23 (24.5%) | 22 (23.4%) | 45 (23.9%) |
|    PhD | 8 (8.5%) | 3 (3.2%) | 11 (5.9%) |
| *Current situation* | | | |
|    Student only | 53 (56.4%) | 56 (59.6%) | 109 (58.0%) |
|    Worker only | 17 (18.1%) | 19 (20.2%) | 36 (19.1%) |
|    Student and worker | 18 (19.1%) | 18 (19.1%) | 36 (19.1%) |
|    Unemployed | 6 (6.4%) | 1 (1.1%) | 7 (3.7%) |
| *VR Use* | | | |
|    Never use | 55 (58.5%) | 55 (58.5%) | 110 (58.5%) |
|    Rarely use | 24 (25.5%) | 22 (23.4%) | 46 (24.5%) |
|    Sometimes use | 9 (9.6%) | 8 (8.5%) | 17 (9.0%) |
|    Often use | 6 (6.4%) | 9 (9.6%) | 15 (8.0%) |
| *Risk preference* | | | |
|    Mean (SD) | 6.915 (1.663) | 6.681 (1.827) | 6.798 (1.746) |

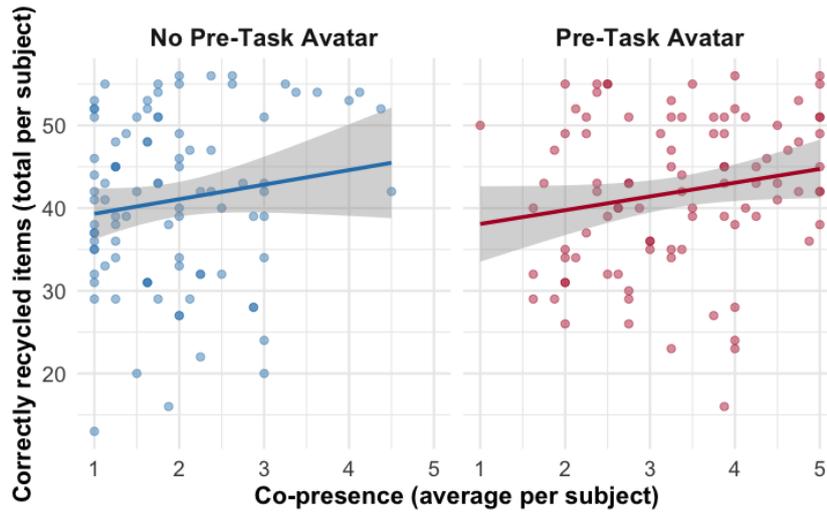

Figure D.2: Total of *Correctly recycled* items (total per subject) and *Co-presence* (per subject) by treatment

The average pair *Co-presence* score across all observations is 2.620 (SD = 0.965). *Pair Co-presence* scores (considering the average per pair) are higher when the threshold is reached (Wilcoxon rank-sum test: $z = -2.493$, $p = 0.012$). When the threshold is not reached, the average *Pair Co-presence* score is in fact only 2.342 (SD = 0.735), compared to 2.887 (SD = 1.090) when the threshold is reached (see Figure D.3).

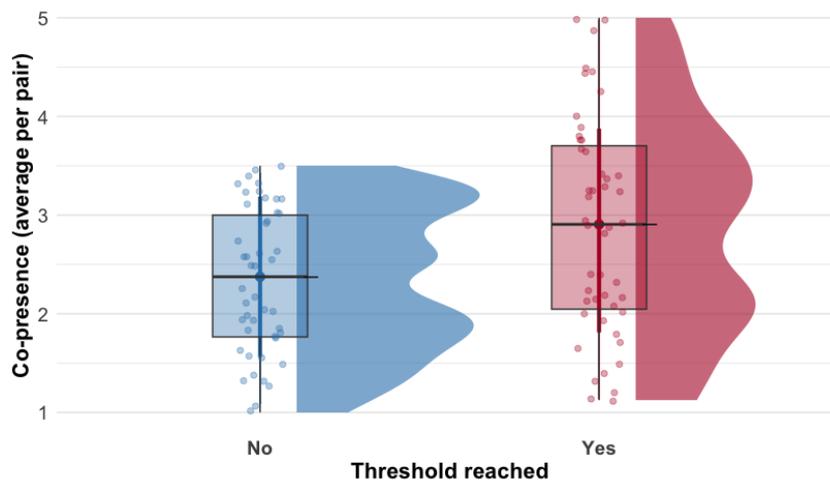

Figure D.3: Co-presence (average per pair) by *Threshold*

### D.3. Preliminary analyses Section 5.3

The average number of *Correctly recycled* items is 40.936 ($SD = 10.208$) in the *No Pre-Task Avatar* treatment and 41.936 ($SD = 9.151$) in the *Pre-Task Avatar* treatment. A Wilcoxon rank-sum test revealed no statistically significant difference between treatments ($z = -0.554$, $p = 0.581$).

Regarding the *Recycling threshold*, 28 pairs in the *Pre-Task Avatar* treatment successfully reached the threshold compared to 20 pairs in the *No Pre-Task Avatar* treatment. Conversely, 19 pairs in the *Pre-Task Avatar* treatment and 27 pairs in the *No Pre-Task Avatar* treatment failed to reach the threshold. A one-tailed Fisher's exact test indicated that this difference is marginally statistically significant ($p = 0.074$).

The *Environmental score*, based on 10 questions about pro-environmental behavior in daily life measured on a 5-point Likert scale, is 4.182 ($SD = 0.451$). In the *No Pre-Task Avatar* treatment, the average score is 4.140 (SD = 0.471), while in the *Pre-Task Avatar* treatment it is 4.224 (SD = 0.429). The difference between the two treatments is not statistically significant according to a Wilcoxon rank-sum test (z = –0.959, p = 0.339). A Spearman correlation test shows no significant relationship between the number of *Correctly recycled* items and the *Environmental score* (Spearman's $\rho = 0.0498$, p = 0.497). Only question 2 (reverse worded) is marginally correlated with the number of correctly recycled items (Spearman's $\rho = -0.140$, p = 0.056), indicating that subjects who reported that recycling requires too much effort tended to recycle fewer items in the VR experiment.

The average score on the *IPQ* questionnaire, measured on a 7-point Likert scale, is 4.706 (SD = 0.715). In the *No Pre-Task Avatar* treatment, the average score is 4.720 (SD = 0.671), while in the *Pre-Task Avatar* treatment it is 4.692 (SD = 0.761). There is not a significant difference of the average IPQ score between the two treatments as shown by a Wilcoxon rank-sum test (z = 0.117, p = 0.908). Furthermore, there is not a significant difference between the average IPQ score and correctly recycled objects ((Spearman's $\rho = -0.028$, $p = 0.701$). However, among the four categories of the IPQ questionnaire (General Presence [*G*], Spatial Presence [*SP*], Involvement [*INV*], and Experienced Realism [*REAL*]), two showed significant correlations with the number of correctly recycled items: *SP* (Spearman's $\rho = 0.197$, $p = 0.007$) and *REAL*, which was negatively correlated (Spearman's $\rho = -0.175$, $p = 0.017$).

Concerning the *WS* adapted questionnaire, which is also measured on a 7-point Likert scale, the average value of the 10 combined items is 5.17 (SD = 0.863). In the *No Pre-Task Avatar* condition, the average score is 5.082 (SD = 0.869), while in the *Pre-Task Avatar* treatment it is 5.251 (SD = 0.853). No significant difference is to be declared of the WS average score between treatments as confirmed by a Wilcoxon rank-sum test (z = -1.144 , p = 0.254). A Spearman test shows a significant correlations between the average WS score and correctly recycled objects (Spearman's $\rho = 0.476$, $p < 0.001$).

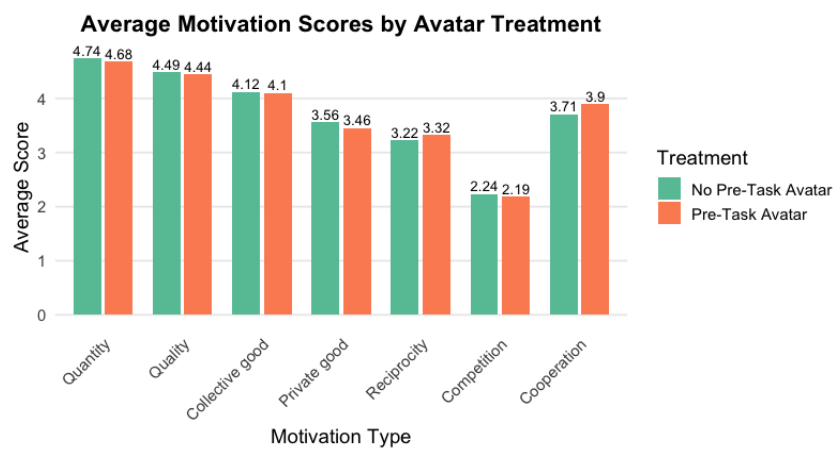

Figure D.4: Recycling motivation questionnaire

Figure D.4 compares the scores obtained in the recycling motivation questionnaire. No significant differences were found between treatments.